\documentclass[superscriptaddress,aps,pra,twocolumn,showpacs,preprintnumbers,amsmath,amssymb,floatfix,10pt]{revtex4-1}

\usepackage{mathrsfs}

\usepackage{graphicx}
\usepackage{dcolumn}
\usepackage{bm}
\usepackage{color}

\def\ba{\begin{eqnarray}}
\def\ea{\end{eqnarray}}
\def\beq{\begin{equation}}
\def\eeq{\end{equation}}

\newcommand{\threej}[6]{\left(\begin{array}{ccc}#1&#2&#3\\#4&#5&#6\end{array}\right)}
\newcommand{\sixj}[6]{\left\{\begin{array}{ccc}#1&#2&#3\\#4&#5&#6\end{array}\right\}}
\newcommand{\ninej}[9]{\left\{\begin{array}{ccc}#1&#2&#3\\#4&#5&#6\\#7&#8&#9\end{array}\right\}}

\usepackage[colorlinks,bookmarks=false,citecolor=blue,linkcolor=red,urlcolor=blue]{hyperref}

\begin{document}

\title{Quantum Magnetism with Polar Alkali Dimers}

\author{Alexey V. Gorshkov}
\affiliation{Institute for Quantum Information, California Institute of Technology, Pasadena, California 91125, USA}
\author{Salvatore R. Manmana}
\affiliation{JILA, NIST, and Department of Physics, University of Colorado, Boulder, CO 80309}
\author{Gang Chen}
\affiliation{JILA, NIST, and Department of Physics, University of Colorado, Boulder, CO 80309}
\author{Eugene Demler}
\affiliation{Physics Department, Harvard University, Cambridge, Massachusetts 02138, USA}
\author{Mikhail D. Lukin}
\affiliation{Physics Department, Harvard University, Cambridge, Massachusetts 02138, USA}
\author{Ana Maria Rey}
\affiliation{JILA, NIST, and Department of Physics, University of Colorado, Boulder, CO 80309}

\date{\today}

\begin{abstract}

We show that dipolar interactions between ultracold polar alkali dimers in optical lattices can be used to realize a highly tunable generalization of the $t$-$J$ model, which we refer to as the $t$-$J$-$V$-$W$ model. The model features long-range spin-spin interactions $J_z$ and $J_\perp$ of XXZ type, long-range density-density interaction $V$, and long-range density-spin interaction $W$, all of which can be controlled in both magnitude and sign independently of each other and of the tunneling $t$. The ``spin" is encoded in the rotational degree of freedom of the molecules, while the interactions are controlled by applied static electric and continuous-wave microwave fields. Furthermore, we show that nuclear spins of the molecules can be used to implement an additional (orbital) degree of freedom that is coupled to the original rotational degree of freedom in a tunable way. The presented system is expected to exhibit exotic physics and to provide insights into strongly correlated phenomena in condensed matter systems. Realistic experimental imperfections are discussed.  

\end{abstract} 




\maketitle

\section{Introduction}

Ultracold diatomic polar molecules have recently attracted a great deal of attention both experimentally and theoretically
\cite{dulieu09,carr09,krems09,baranov08,lahaye09,trefzger11}. Two features of diatomic polar molecules make them particularly interesting as compared to the more typical systems of ultracold alkali atoms. First, polar molecules possess a permanent dipole moment, which can be manipulated with external fields and which can lead to long-range anisotropic interactions. This contrasts with atoms whose interactions are typically short-range and isotropic. Second, the internal level structure of diatomic polar molecules is much richer than that of atoms and, although more difficult to control, allows, in principle, for richer physics. These two features make diatomic polar molecules attractive for numerous applications including quantum computation, quantum simulation, precision measurements, and controlled quantum chemistry \cite{dulieu09,carr09,krems09,baranov08,lahaye09}. 

In Ref.\  \cite{gorshkov11}, it was shown that these two unique features allow ultracold polar molecules in optical lattices to simulate a highly tunable generalization of the $t$-$J$ model \cite{auerbach94,lee06b,ogata08} referred to as the $t$-$J$-$V$-$W$ model. 
This model makes use of the rotational degree of freedom of the molecules and features tunneling $t$, density-density interaction $V$, density-spin interaction $W$, and spin-spin interactions $J_z$ and $J_\perp$ of XXZ type. As a first step towards understanding this model, Ref.\ \cite{gorshkov11} showed that the simplest experimentally realizable  case of the $t$-$J$-$V$-$W$ model with $V = W = J_z = 0$ allows to strongly enhance the 
superconducting (i.e.\ superfluid for our neutral system) region of the 1D phase diagram relative to the usual $t$-$J$ model. 

In the present paper, we provide the details behind the derivation of the $t$-$J$-$V$-$W$ model. In particular, we show that the manipulation of the rotational degree of freedom of the molecules via DC electric and microwave fields allows to achieve full control of the coefficients of the $t$-$J$-$V$-$W$ Hamiltonian and discuss the implications of this control on the accessible manybody physics. Specifically, one can tune the system into exhibiting the physics very similar to the original $t$-$J$ model, whose phase diagram is still highly controversial beyond one dimension  \cite{auerbach94,lee06b,ogata08}. Alternatively, one can access a wide range of other regimes that include the  spin-1/2 XXZ magnet and numerous extensions of the $t$-$J$ model, some of which are believed to exhibit enhanced 
superfluid correlations. We also show how to control the spatial anisotropy of the Hamiltonian by changing the direction of the applied DC electric field and how to control the optical potential experienced by different rotational states by an appropriate choice of lattice laser beams.

Furthermore, we study in detail the generalization of the $t$-$J$-$V$-$W$ model to the case where not only the rotational degree of freedom of the molecules, but also their nuclear degrees of freedom play an important role. Due to the relative simplicity of their production, the only ultracod polar molecules currently available in their electronic, vibrational, and rotational ground states are alkali dimers KRb  \cite{ni08,ni10,demiranda10,aikawa10} and LiCs \cite{deiglmayr08,deiglmayr10}. 
Therefore, we focus on the hyperfine structure of alkali dimers, which has  recently been studied theoretically \cite{aldegunde08,aldegunde09,aldegunde09c,ran10,wall10} and experimentally \cite{ospelkaus10,ospelkaus10b,danzl10}. Specifically, we show how the applied DC electric field can be used to couple and decouple rotational and nuclear degrees of freedom, thus allowing for the control of nuclear spin effects. In the case where nuclear spins are coupled to the rotational degree of freedom, we show that the nuclear spins can function either as classical -- possibly spatially dependent -- magnetic field or as a separate (orbital) quantum degree of freedom with a highly tunable interaction with the rotor. We also point out possible promising applications of the system to quantum information processing. Since ultracold ground-state polar alkali dimers are already available in experiments \cite{ni08,ni10,demiranda10,aikawa10,deiglmayr08,deiglmayr10} and are even loaded in optical lattices \cite{demiranda10}, we expect our results to be immediately applicable to current experiments.

Our work builds on an extensive body of literature studying the many-body dynamics of polar molecules in a lattice and making use of 
the internal  rotational structure \cite{barnett06,micheli06,brennen07,buchler07b,watanabe09,wall09,yu09b,krems09,wall10,schachenmayer10,perezrios10,trefzger10,herrera10,kestner11} 
and of fine and   hyperfine structure of molecules with a single electron outside a closed shell \cite{micheli06,brennen07,perezrios10}. We would also like to specifically highlight recent work in Refs.\ \cite{wall10,kestner11}, which make important steps towards the understanding of the effects of hyperfine structure on many-body physics with alkali dimers. 

The remainder of the paper is organized as follows. In Sec.\ \ref{sec:ham}, we introduce the $t$-$J$-$V$-$W$ Hamiltonian in the presence of both a rotational and a nuclear degree of freedom and describe its main features. Then, in Secs.\ \ref{sec:hf} - \ref{sec:stab}, we present  
a detailed derivation and discussion of this Hamiltonian. In particular, in Sec.\ \ref{sec:hf}, we study the rotational and hyperfine structure of the molecules in the presence of a DC electric field. In Sec.\ \ref{sec:opt}, we study the optical potential and the associated tensor shifts. In Sec.\ \ref{sec:deriv}, we use the results of Secs.\ \ref{sec:hf} and \ref{sec:opt} to give a detailed derivation of the final Hamiltonian. In Sec.\ \ref{sec:stab}, we find the regimes, in which the model is stable to loss via chemical reactions. Finally, in Sec.\ \ref{sec:conc}, we present the conclusions. Appendix \ref{sec:matel} presents formulas useful for studying the single-molecule Hamiltonian and dipole-dipole interactions between molecules. Appendix \ref{sec:iat} describes the phenomenon of interaction-assisted tunneling, which arises if one considers small corrections to the $t$-$J$-$V$-$W$ model.





\section{The Hamiltonian and its features \label{sec:ham}}


In this Section, we introduce the $t$-$J$-$V$-$W$ Hamiltonian in the presence of both a rotational and a nuclear degree of freedom and describe its main features. The detailed derivation is postponed until Secs.\ \ref{sec:hf}-\ref{sec:stab}.

We consider diatomic polar molecules confined to a single  plane 
(e.g.\ using a strong 1D optical lattice) and subject to a DC electric  and, possibly, one or more continuous-wave (CW) microwave fields. Furthermore, in that plane, the molecules are assumed to be loaded in the lowest band of a 2D optical lattice. Such a system is not far out of reach experimentally: indeed, loading of KRb molecules into 1D \cite{demiranda10} and 3D \footnote{Private communication with Jun Ye.} lattices and of homonuclear Cs$_2$ molecules into a 3D lattice \cite{danzl10} has already been demonstrated.

As shown in Secs.\ \ref{sec:hf}-\ref{sec:deriv}, taking into account the applied DC and microwave fields, we can reduce  the internal structure of  each molecule to  a tensor product of a two-level dressed rotational degree of freedom (dressed states labeled by $|m_0\rangle$ and $|m_1\rangle$; angular momentum operator on site $j$ labeled by $\mathbf{S}_j$) and a two-level nuclear degree of freedom (states labeled by $|\uparrow\rangle$ and $|\downarrow\rangle$;  angular momentum operator on site $j$ labeled by $\mathbf{T}_j$). In Secs.\ \ref{sec:hf}-\ref{sec:stab}, we derive the following Hamiltonian: 
\begin{eqnarray} \label{eq:nuctun}
H &\!=\!& - \sum_{\langle i,j\rangle m \sigma } t_m \left[c^\dagger_{i m \sigma} c_{j m \sigma} + \textrm{h.c.}\right] \nonumber \\
&& + \frac{1}{2} \sum_{i \neq j} V_\textrm{dd}(\mathbf{R}_i\!-\!\mathbf{R}_j)  \Bigg[ J_z S^z_{i} S^z_{j}+ \frac{J_\perp}{2} (S^+_i S^-_j + S^-_i S^+_j)  \nonumber \\
&&  + V  n_i n_j + W (n_i S^z_{j} + n_j S^z_{i}) \Bigg]+ A \sum_{i} S^z_i T^z_i.
\end{eqnarray}
%
%
%
%
%
This Hamiltonian, together with the full control over its coefficients and with the detailed study of the hyperfine structure, is the main result of the present paper. The first term ($\propto t_m$) describes tunneling of molecules; the second term ($\propto V_\textrm{dd}$) describes dipole-dipole interactions; while the last term ($\propto A$) describes hyperfine interactions. Let us describe each of these terms,  including the necessary definitions and the physical origin.

Let us begin with the tunneling term $\propto t_m$. The bosonic or fermionic creation operator $c^\dagger_{j m \sigma}$ creates a molecule on site $j$ in the dressed rotor state $m$  ($= m_0$ or $m_1$) and nuclear state $\sigma$ ($= \uparrow$ or $\downarrow$). 
The notation $\langle i,j\rangle$ indicates the sum over nearest neighbors, where each pair of nearest neighbors is included only once. Throughout the paper, we set $\hbar = 1$.
As we will show in Sec.\ \ref{sec:opt}, the tunneling amplitudes $t_{m_0}$ and $t_{m_1}$ can be made either equal or different by choosing the polarization and frequency of the optical fields creating the lattice and by choosing the dressed states $|m_0\rangle$ and $|m_1\rangle$. The overall magnitude of the amplitudes $t_m$ can be tuned from zero up to a few 
kHz by changing the intensity of the optical fields. Notice that the chemical potential is not included in Eq.\ (\ref{eq:nuctun}) since $H$ conserves the total number of molecules in each internal state $|m \sigma\rangle$ and since the system is not coupled to a reservoir of molecules. The effect of a chemical potential can be modeled by controlling the total number of molecules in each internal state $|m \sigma \rangle$ during the preparation stage.

Let us now describe the dipole-dipole interaction term $\propto V_\textrm{dd}$. The operator $n_{j m \sigma} = c^\dagger_{j m \sigma} c_{j m \sigma}$ counts the number of molecules on site $j$ in the dressed rotor state $m$ and nuclear state $\sigma$, while the operator $n_{j m} = \sum_{\sigma} n_{j m \sigma}$ counts the number of molecules on site $j$ in the dressed rotor state $m$ irrespective of the nuclear state.  We assume [see Sec.\ \ref{sec:stab}] that on-site interactions and/or on-site decay for two molecules are so large that molecules obey the hardcore constraint, i.e.\ each site can be occupied by either 0 or 1 molecules [although it is straightforward to extend the model to finite on-site interactions (see e.g.\ Ref.\ \cite{barnett06})]. The operators $S_j^z = (n_{j m_0}-n_{j m_1})/2$, $S_j^+ = \sum_{\sigma} c^\dagger_{j m_0 \sigma} c_{j m_1 \sigma}$, and $S_j^- = (S_j^+)^\dagger$ are the usual spin-1/2 angular momentum operators on site $j$ describing the two-level dressed rotor degree freedom and satisfying $[S_j^z,S_j^\pm] = \pm S_j^\pm$.  

\begin{figure}[t]
\begin{center}
\includegraphics[width = 0.8 \columnwidth]{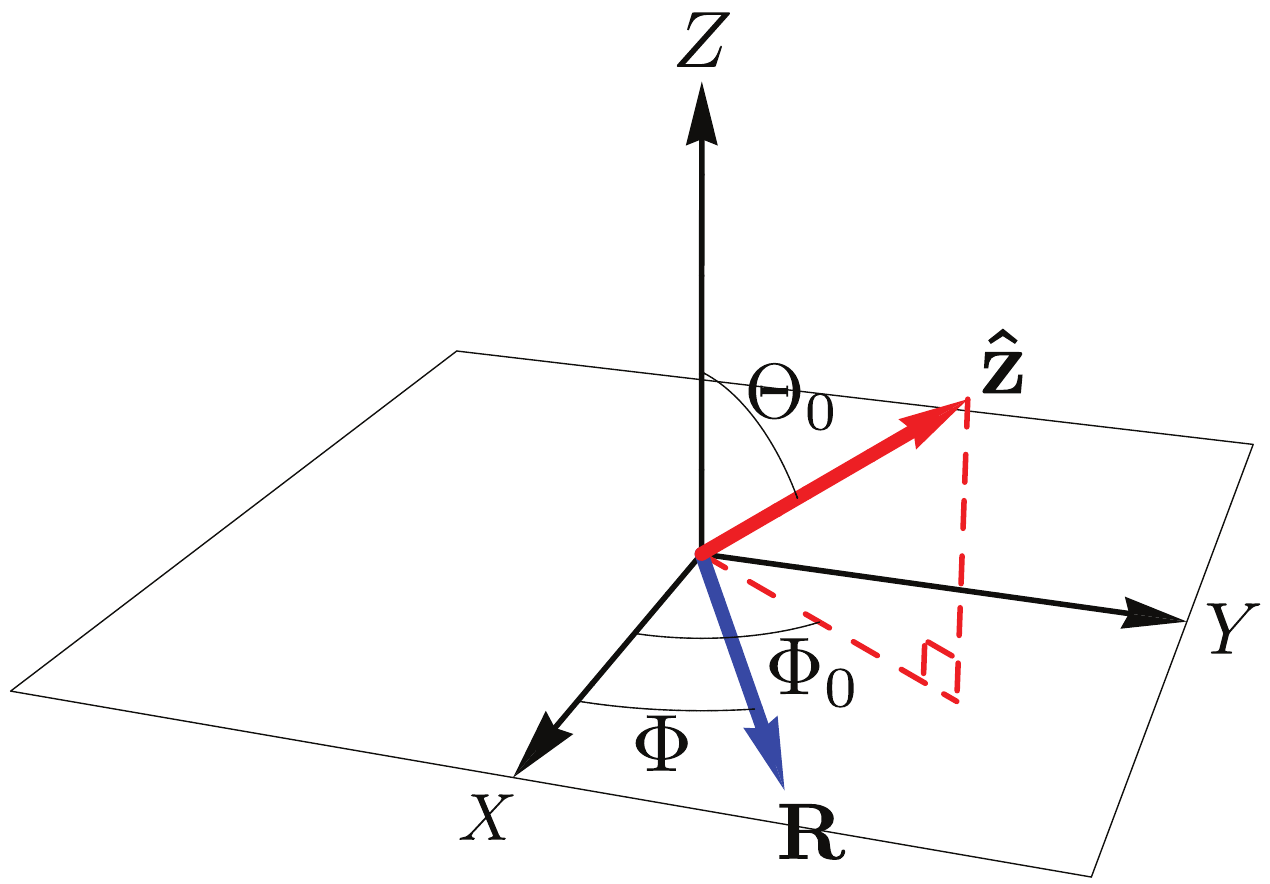}
\caption{(color online). The geometry of the setup. The molecules are assumed to be in the $X$-$Y$ plane. A typical vector $\mathbf{R}$ in that plane has polar coordinates $(R,\Phi)$. The direction of the DC electric field has spherical coordinates $(\Theta_0,\Phi_0)$ in the $X$-$Y$-$Z$ coordinate system. The quantization axis $\mathbf{\hat{z}}$ for the spins lies along the applied DC electric field. The other two axes ($\mathbf{\hat{x}}$ and $\mathbf{\hat{y}}$) of the spin coordinate system are not shown. The cosine of the angle between $\mathbf{R}$ and $\mathbf{\hat{z}}$ is equal to $\hat{\mathbf{R}} \cdot \hat{\mathbf{z}} = \sin \Theta_0 \cos(\Phi - \Phi_0)$ [see Eq.\ (\ref{eq:vdd})]. \label{fig:angles}}
\end{center}
\end{figure}

As shown in Fig.\ \ref{fig:angles}, the 2D plane, which the molecules are confined to, is assumed to be the $X$-$Y$ plane, while the vector perpendicular to it defines the $Z$-axis (note the use of the upper case to denote the spatial axes). All angular momenta are, on the other hand, quantized along the $z$-axis (note the use of the lower case to denote angular momentum axes), which is the axis along which the DC electric field is applied. The $z$-axis has spherical coordinates $(\Theta_0, \Phi_0)$ relative to the $X$-$Y$-$Z$ coordinate system. In Eq.\ (\ref{eq:nuctun}), $\mathbf{R}_i$ is the position of site $i$ in the $X$-$Y$ plane. Classical dipole-dipole interaction energy between two unit electric dipoles  oriented along $\hat{\mathbf{z}}$ and located at sites $i$ and $j$ is then given by
\begin{eqnarray} \label{eq:vdd}
V_\textrm{dd}(\mathbf{R}) &=& \frac{1}{4 \pi \epsilon_0 R^3} \left[1 - 3 (\hat{\mathbf{R}} \cdot  \hat{\mathbf{z}})^2\right] \nonumber \\
&=& \frac{1}{4 \pi \epsilon_0 R^3} \left[1-3 \sin^2 \Theta_0 \cos^2 (\Phi - \Phi_0)\right],
\end{eqnarray} 
where $\mathbf{R} = \mathbf{R}_i - \mathbf{R}_j = (R,\Phi)$ in polar coordinates, and $\hat{\mathbf{R}} = \mathbf{R}/R$ is a unit vector along $\mathbf{R}$. 
In Eq.\ (\ref{eq:vdd}), $\hat{\mathbf{R}} \cdot  \hat{\mathbf{z}}  = \sin \Theta_0 \cos(\Phi - \Phi_0)$ is the cosine of the angle between $\mathbf{R}$ and $\hat{\mathbf{z}}$. $V_\textrm{dd}(\mathbf{R})$ is used in Eq.\ (\ref{eq:nuctun}) and reproduces the usual dipole-dipole interaction behavior with head-to-tail attraction when $\hat{\mathbf{z}} = \hat{\mathbf{R}}$ and side-to-side repulsion when $\hat{\mathbf{z}} = \hat{\mathbf{Z}}$. The dipole-dipole interaction term in Eq.\ (\ref{eq:nuctun}) is multiplied by 1/2 since we double-count.

The origin of the $J_z$, $J_\perp$, $V$, and $W$ terms in Eq.\ (\ref{eq:nuctun}) can be evinced with the following simple example, which does not involve the application of microwave fields. The rotational degree of freedom of a single molecule is described by the angular momentum operator $\mathbf{N}$. Let us pick as $|m_0\rangle$ and $|m_1\rangle$ the lowest two $N_z = 0$ states of the molecule in the presence of a DC electric field along $\hat{\mathbf{z}}$. Due to the applied electric field, these states are not eigenstates of $\mathbf{N}^2$ and possess nonzero dipole moments. One can then intuitively think of the ground state $|m_0\rangle$ as a dipole $\boldsymbol{\mu}_0 = \mu_0 \hat{\mathbf{z}}$  oriented along the DC field (i.e.\ $\mu_0 > 0$) and of the excited state $|m_1\rangle$ as a dipole $\boldsymbol{\mu}_1 = \mu_1 \hat{\mathbf{z}}$ oriented against the DC electric field (i.e.\ $\mu_1 < 0$). Let us now consider classical dipole-dipole interaction energy $E_{dd}$ between a dipole $\boldsymbol{\mu}_i = (\mu_0 n_{i m_0} + \mu_1 n_{i m_1}) \hat{\mathbf{z}}$ at site $i$ and a dipole $\boldsymbol{\mu}_j = (\mu_0 n_{j m_0} + \mu_1 n_{j m_1}) \hat{\mathbf{z}}$ at site $j$, where $n_{k m}$ indicates whether the molecule on site $k$ is in state $|m\rangle$ ($n_{k m} = 1$) or not ($n_{k m} = 0$):
\ba
&&E_{dd} = \frac{1}{4 \pi \epsilon_0 |\mathbf{R_i} - \mathbf{R_j}|^3} \left[\boldsymbol{\mu}_i \cdot \boldsymbol{\mu}_j - 3 (\boldsymbol{\mu}_i \cdot \hat{\mathbf{z}}) (\boldsymbol{\mu}_j \cdot \hat{\mathbf{z}})\right] \label{eq:edd}  \\
&& = V_{dd} (\mathbf{R_i} - \mathbf{R_j}) (\mu_0 n_{i m_0} + \mu_1 n_{i m_1}) (\mu_0 n_{j m_0} + \mu_1 n_{j m_1}) \nonumber \\
&& = V_{dd} (\mathbf{R_i} - \mathbf{R_j}) \left[ J_z S_i^z S_j^z + V n_i n_j + W(n_i S_j^z + n_j S_i^z)\right], \nonumber
\ea
where  $J_z = (\mu_0 - \mu_1)^2$, $V = (\mu_0 + \mu_1)^2/4$, and $W = (\mu_0^2 - \mu_1^2)/2$. 
The $V$ term describes density-density interactions, and is the only term that survives if one averages $E_{dd}$ over the internal states $|m_0\rangle$ and $|m_1\rangle$ of each of the two molecules. Furthermore, only the $V$ term survives if $\mu_0 = \mu_1$, in which case dipole-dipole interaction cannot depend on the internal states of the two molecules. The $J_z$ term describes an Ising-type spin-spin interaction. 
Since $J_z$ is non-negative in this example, it favors, for $V_{dd} > 0$, antialignment of molecules on sites $i$ and $j$. This makes sense since two side-by-side dipoles repel if they are aligned, but attract if they are antialigned. Finally, the $W$ term describes spin-density interaction. In the language of quantum magnetism, 
the presence of a molecule on site $i$ creates, via the term $W n_i S_j^z$, an effective magnetic field along $\hat{\mathbf{z}}$ for the spin on site $j$. 
As one can see from this discussion, an important difference of our Hamiltonian from Refs.\ \cite{micheli06, brennen07}, which also 
engineer magnetic models using molecules in optical lattices,  
is that we use dipole-dipole interactions in first order (rather than second order), which allows for stronger interactions. 
 
To understand the origin of the $J_\perp$ term in Eq.\ (\ref{eq:nuctun}), one has to take into account the transition dipole moment $\mu_{01}$ between $|m_0\rangle$ and $|m_1\rangle$ \cite{barnett06}. In the same way, in which an optical excitation can be exchanged between two two-level atoms that are within an optical wavelength of each other \cite{dicke54}, 
the $J_\perp$ term describes the exchange of a microwave excitation; in this example, $J_\perp = 2 \mu_{01}^2$. For a molecule on site $i$ and a molecule on site $j$ that share one  microwave excitation, the $J_\perp$ term is diagonalized by the symmetric and antisymmetric states $(|m_0 m_1\rangle_{i j} \pm |m_1 m_0\rangle_{ij})/\sqrt{2}$, which would be the microwave equivalent of optical superradiant and subradiant states \cite{dicke54}.  The presence of the $J_\perp$ term is one of the main differences of Eq.\ (\ref{eq:nuctun}) from the Hamiltonian discussed in Ref.\ \cite{wall10}. While this simple example illustrates the physical origin of the dipole-dipole interaction terms featured in Eq.\ (\ref{eq:nuctun}), we will show in Sec.\ \ref{sec:deriv2}, that this form of dipole-dipole interactions is much more general. In particular, we will show that it applies even when microwave fields are applied and when states with $N_z \neq 0$ are involved.

The recipe for controlling dipole-dipole interactions with applied DC electric and microwave fields is one of the main results of the present paper. Specifically, as we will discuss in Sec.\ \ref{sec:magn}, the spatial anisotropy of the interactions can be controlled via $(\Theta_0,\Phi_0)$. More importantly, as we will discuss in Secs.\ \ref{sec:magn}, \ref{sec:tjv} and derive in Sec.\ \ref{sec:deriv2}, by tuning the strength of the DC electric field as well as the frequency and intensity of the applied microwave field(s), one can achieve 
complete control over signs and relative amplitudes of the coefficients $V$, $W$, $J_z$, and $J_\perp$. The strength of the resulting dipole-dipole interactions (quoted for nearest neighbors separated by $500$ nm) is $\sim 0.4$ kHz in KRb and $\sim 40$ kHz in LiCs. These dipole-dipole interactions (particularly in the case of LiCs) are substantially stronger than superexchange interactions in cold atoms ($ \ll 1$ kHz \cite{trotzky08}), making magnetism easier to access in such molecular systems than in atomic systems. 

Finally, let us describe the hyperfine interaction term ($\propto A$) in Eq.\ (\ref{eq:nuctun}). The operator $n_{j \sigma} = \sum_{m} n_{j m \sigma}$ counts the number of molecules on site $j$ with nuclear spin $\sigma$ irrespective of the rotational state. The operators $T_j^z = (n_{j \uparrow} - n_{j \downarrow})/2$, $T_j^+ = \sum_m c^\dagger_{j m \uparrow} c_{j m \downarrow}$, and $T_j^- = (T_j^+)^\dagger$ are the usual spin-1/2 angular momentum operators on site $j$ describing the two-level nuclear degree freedom and satisfying $[T_j^z,T_j^\pm] = \pm T_j^\pm$. The hyperfine interaction of the form $A S^z_i T^z_i$ relies on the fact [see Sec.\ \ref{sec:hf}] that, for a generic DC electric field, the hyperfine interaction can be projected on states $|m_0\rangle$ and $|m_1\rangle$ and, moreover, is diagonal in the same nuclear spin basis in both states (the basis, in which the two nuclei are decoupled from each other). Thus $A S^z_i T^z_i$ simply reflects the fact that the energy difference between any two of these eigenstates ($|\uparrow\rangle$ and $|\downarrow\rangle$) is generally not the same in $|m_0\rangle$ and $|m_1\rangle$: the flip of the nuclear degree of freedom from $|\downarrow\rangle$ do $|\uparrow\rangle$ in $|m_0\rangle$ takes an energy larger by an amount $A$ than in $|m_1\rangle$.   In Sec.\ 
 \ref{sec:deriv1}, we show that the hyperfine interaction constant $A$ can be tuned, via the strength of the DC electric field and via the choice of nuclear spin states, from zero to almost any value up to $\sim 1$ MHz in KRb and up to $\sim 100$ kHz in LiCs. 
Moreover, as we will note in Secs.\ \ref{sec:nuc} and \ref{sec:deriv1}, while the interaction $S^z_i T^z_i$ is the easiest form of the hyperfine interaction that one can obtain, any interaction between $\mathbf{S_i}$ and $\mathbf{T_i}$ is, in principle, achievable.

Both fermionic (${}^{40}$K${}^{87}$Rb \cite{ni08}) and bosonic (${}^7$Li${}^{133}$Cs \cite{deiglmayr08} and ${}^{41}$K${}^{87}$Rb \cite{aikawa10}) species are available experimentally. The bosonic \cite{goral02, dallatorre06,yi07, menotti07, danshita09,pollet10,capogrosso-sansone10,burnell09,barnett06,wall09, wall10,yu09b, lin10b,schachenmayer10,dalmonte11} and fermionic \cite{watanabe09,wall09, wall10,yu09b, lin10b,schachenmayer10,kestner11, mikelsons11,he11} cases are expected to give rise to different physics.  


If tunneling in the third direction is negligible or if stabilization against collapse and/or chemical reactions in the third direction can be achieved without strong dipole-dipole repulsion (see Sec.\ \ref{sec:stab}), we can extend the Hamiltonian to 3D. A 1D geometry \cite{gorshkov11}, 
as well as non-square lattices can also be considered.

\subsection{Rotational degree of freedom alone \label{sec:rot}}

In this Section, we ignore the nuclear degree of freedom in Eq.\ (\ref{eq:nuctun}) and discuss the tunability of the resulting model, as well as the physics that can be accessed with it.

\subsubsection{Quantum magnetism \label{sec:magn}}

In this Section, we further suppose that the tunneling is negligible. The simplest scenario is then the case of a  single molecule per site. In this case, $n_i = 1$ for all sites $i$. This means that the term in the Hamiltonian proportional to $V$ is a constant and can be dropped. The term proportional to $W$ gives an effective magnetic field on each site. 
Ignoring edge effects, this magnetic field is uniform, making the $W$ term commute with the Hamiltonian. In this case, the $W$ term can also be ignored, so that Eq.\ (\ref{eq:nuctun}) reduces to
 \begin{eqnarray} 
H \!=\! \frac{1}{2} \! \sum_{i \neq j} \! V_\textrm{dd}(\mathbf{R}_i\!-\!\mathbf{R}_j)  \! \left[J_z  S^z_i S^z_j \!+\! \frac{J_\perp}{2} (S^+_i S^-_j \!+\! S^-_i S^+_j)  \right]\!. \label{eq:simple}
\end{eqnarray}
The important features of the interaction in Eq.\ (\ref{eq:simple}) are that it is long-range, anisotropic in both space 
and spin, and highly tunable via the magnitude of the DC electric field, $\Theta_0$, $\Phi_0$, the choice of rotational states, and the number, frequency, and intensity of applied microwave fields. 
In Ref.\ \cite{hauke10} and Refs.\ \cite{yu09b,schachenmayer10}, this Hamiltonian is studied in the 1D geometry in the context of ions and molecules, respectively.  The $J_z = 0$ case is also studied in the context of molecules in Ref.\ \cite{barnett06}. In Ref.\ \cite{herrera10}, this Hamiltonian is studied in the context of exciton-impurity interactions generated with polar molecules. Related lattice models with dipolar interactions are also studied in the context of Frenkel excitatons  \cite{agranovich09,zoubi05}.   In Ref.\ \cite{rabl07}, a similar Hamiltonian is studied in the context of molecular Wigner crystals for quantum memory applications.

As we will show in Sec.\ \ref{sec:deriv2}, if we parametrize $J_z$ and $J_\perp$ as $J_z = |J| \cos \psi$ and $J_\perp = |J| \sin \psi $, any value of $\psi$ can be achieved by an appropriate combination of DC electric and microwave fields. In other words, one can access the full parameter space. For example, one can get a classical Ising model with $J_\perp = 0$, a pure XX model with $J_z = 0$, or the SU(2)-symmetric Heisenberg interaction ($J_z = J_\perp$).

By changing the direction $(\Theta_0,\Phi_0)$ of the applied electric field, one can control the spatial anisotropy of the interaction \cite{barnett06,ticknor11,cremon10}. In particular, one can  set to zero couplings along one or two directions in the $X$-$Y$ plane. We assume a 2D square-lattice geometry and define $V_{X,Y} = [1 - 3  \sin^2 \Theta_0 \cos^2 (\Phi_{X,Y} - \Phi_{0})] (X^2 + Y^2)^{-3/2}$ as the coupling coefficient between the origin and the site with coordinates $(X,Y)$. The coordinates are given in units of lattice spacing $a = \lambda/2$, where $\lambda$ is the wavelength of the light used to form the lattice. 
Here $\Phi_{X,Y} = \textrm{Arg}(X + i Y)$ is the polar angle of the vector $(X,Y)$ in the plane, and $(\Theta_0,\Phi_0)$ are the polar and  azimuthal angles of the applied DC electric field in the $(X,Y,Z)$ coordinate system.

\begin{figure}[t!]
\begin{center}
\includegraphics[width = 0.99 \columnwidth]{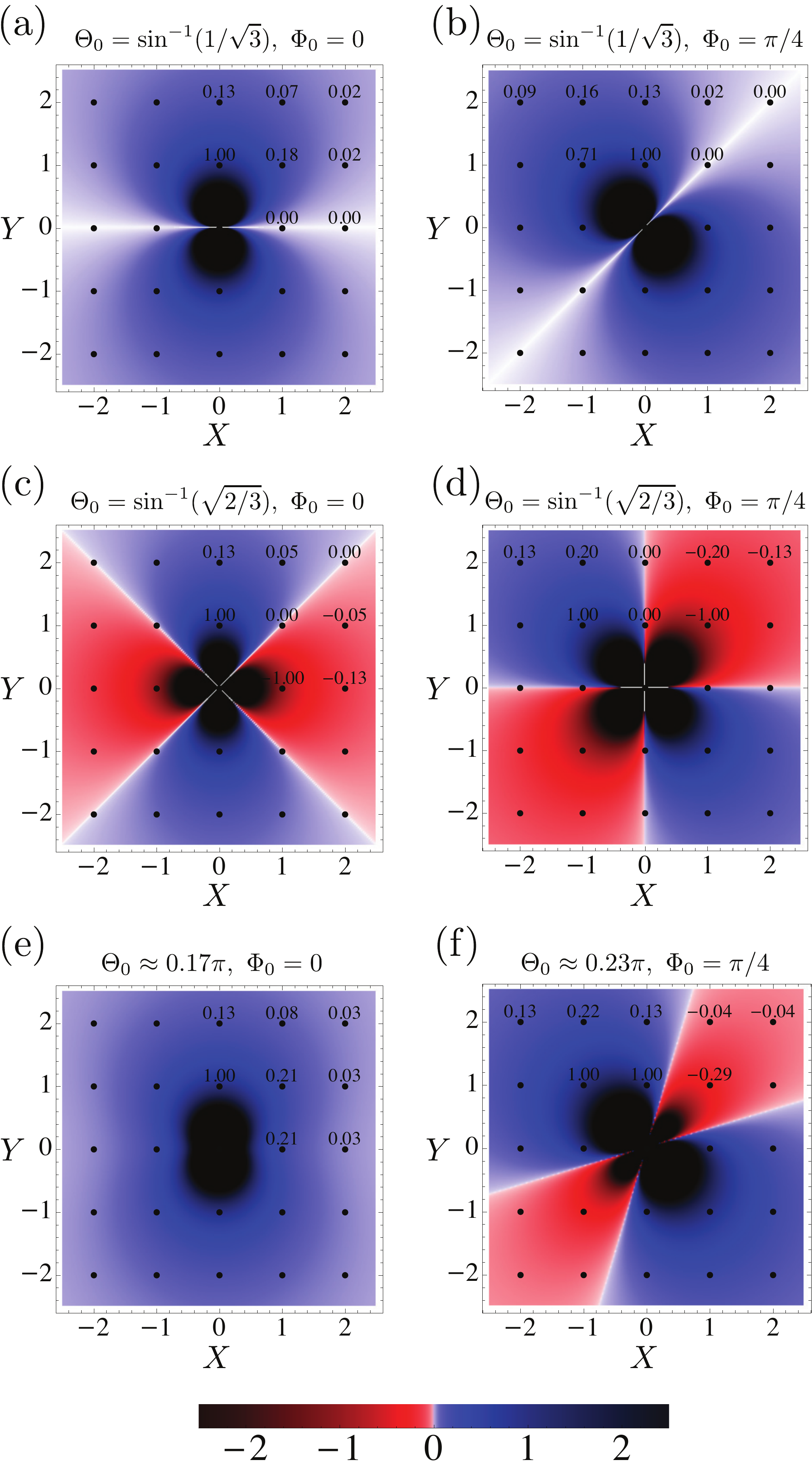}
\caption{(color online). Control over the spatial anisotropy of the interactions. The numbers near particular sites show the dipole-dipole interaction coefficient $V_{X,Y}$ (normalized by the magnitude of the largest $V_{X,Y}$) between the origin and the site $(X,Y)$.  (a) $\Theta_0 = \sin^{-1}(1/\sqrt{3}) \approx 0.20 \pi$, $\Phi_0 = 0$. (b) $\Theta_0 = \sin^{-1}(1/\sqrt{3})$, $\Phi_0 = \pi/4$. (c) $\Theta_0 = \sin^{-1}(\sqrt{2/3}) \approx 0.30 \pi$, $\Phi_0 = 0$. (d) $\Theta_0 = \sin^{-1}(\sqrt{2/3})$, $\Phi_0 = \pi/4$. (e) $\Theta_0 =  \sin^{-1}(\sqrt{(2 \sqrt{2}-1)/(6 \sqrt{2}- 3/2)}) \approx 0.17 \pi$, 
$\Phi_0 = 0$. (f) $\Theta_0 =  \sin^{-1}(\sqrt{(2-1/ \sqrt{2})/3)}) \approx 0.23 \pi$, 
$\Phi_0 = \pi/4$. \label{fig:anisot}} 
\end{center}
\end{figure}

In three dimensions, two cones making an angle $\cos^{-1}(1/\sqrt{3}) \approx 0.30 \pi$ 
with the applied DC electric field (which points along $\mathbf{\hat{z}}$) give vanishing dipole-dipole interactions.  As we tilt the electric field from $\mathbf{\hat{Z}}$ towards the $X$-$Y$ plane (i.e.\ increase $\Theta_0$), interactions in the plane start changing magnitude in an anisotropic fashion. In particular, when $\sin \Theta_0 = 1/\sqrt{3}$ ($\Theta_0 \approx 0.20 \pi$), 
the cone of vanishing interaction touches the $X$-$Y$ plane giving one line in the plane along which dipole-dipole interactions vanish. As shown in Figs.\ \ref{fig:anisot}(a) and (b), 
using $\Phi_0 = 0$ or $\Phi_0 = \pi/4$, we can set $V_{1,0} = 0$ or $V_{1,1} = 0$, respectively.  An interesting feature of setting $V_{1,1} = 0$ is that (provided interactions beyond $V_{1,-1}$ are ignored), this turns a square lattice into an effective triangular lattice. We note that the interactions $V_{X,Y}$ in Fig.\ \ref{fig:anisot} are normalized by the magnitude of the largest one.

As we tilt the electric field further, the single line of vanishing interaction splits into two, and the angle between the two lines increases up to a maximum of $2 \cos^{-1}(1/\sqrt{3})\approx 0.61 \pi$ 
when $\mathbf{\hat{z}}$ (i.e.\ the electric field) is in the $X$-$Y$ plane.  This way, for example, coupling along two orthogonal directions can be set to zero when $\sin \Theta_0 = \sqrt{2/3}$. In particular, as shown in Fig.\ \ref{fig:anisot}(c),
at $\Phi_0 = 0$, we have $V_{1,1} = V_{1,-1} = 0$, while $V_{0,1}$ is unchanged and while the sign of $V_{1,0}$ flips. Alternatively, as shown in Fig.\ \ref{fig:anisot}(d), at $\Phi_0 = \pi/4$, we get $V_{1,0} = V_{0,1} = 0$, while  $V_{1,-1}$ is unchanged and while the sign of $V_{1,1}$ is flipped. 

Finally, some couplings can be set equal to each other. For example [Fig.\ \ref{fig:anisot}(e)], at $\sin \Theta_0 \approx 0.51$ and $\Phi_0 = 0$, we have $V_{1,0} = V_{1,1} \approx 0.2 V_{0,1}$. Alternatively  [Fig.\ \ref{fig:anisot}(f)], at $\sin \Theta_0 = 0.66$ and $\Phi_0 = \pi/4$, we have $0.29 V_{0,1} = 0.29 V_{-1,1} = - V_{1,1}$. 

In experiments, it is difficult to achieve a perfect occupation of exactly one molecule per site. There will, thus, always be empty sites, which will play the role of defects in the corresponding spin model. Furthermore, empty sites can be introduced on purpose to emulate the effect of static non-magnetic impurities in quantum magnets.

\subsubsection{The $t$-$J$-$V$-$W$ model \label{sec:tjv}}

Allowing for tunneling between sites, we arrive at the following Hamiltonian, which we refer to as the $t$-$J$-$V$-$W$ model:
\begin{eqnarray} \label{eq:tun}
H &\!=\!& - \sum_{\langle i,j\rangle m \sigma } t_m \left[c^\dagger_{i m \sigma} c_{j m \sigma} + \textrm{h.c.}\right] \nonumber \\
&& + \frac{1}{2} \sum_{i \neq j} V_\textrm{dd}(\mathbf{R}_i\!-\!\mathbf{R}_j)  \Bigg[ J_z S^z_{i} S^z_{j}+ \frac{J_\perp}{2} (S^+_i S^-_j + S^-_i S^+_j)  \nonumber \\
&&  + V  n_i n_j + W (n_i S^z_{j} + n_j S^z_{i}) \Bigg]. 
\end{eqnarray}
This model is an extension of the $t$-$J$ model \cite{troyer93,dagotto94,moreno11}. The $t$-$J$ model emerges from the large-$U$ expansion of the Hubbard model. Despite significant efforts to identify the phase diagram of the $t$-$J$ model, only the 1D phase diagram is relatively well-established (via numerical methods) \cite{troyer93,dagotto94,moreno11}. 
It has also been demonstrated  that, in 1D, the addition of repulsive nearest-neighbor interactions $V \sum_i n_i n_{i+1}$ (giving rise to the so-called $t$-$J$-$V$ model) and next-nearest-neighbor repulsive interactions $V' \sum_i n_i n_{i+2}$ (giving rise to the so-called $t$-$J$-$V$-$V'$ model) can strengthen superconducting (i.e.\ superfluid for our neutral system) correlations 
in the $t$-$J$ model \cite{troyer93}. Ref.\ \cite{troyer93} also argues that this effect will manifest itself 
in a 2D geometry as well. A confirmation of this statement can have important implications in the understanding of high-temperature superconductivity in cuprates.

The highly tunable model in Eq.\ (\ref{eq:tun}) provides unique opportunites to study a generalized $t$-$J$ model in 1D and 2D geometries. Some of the important features of Eq.\ (\ref{eq:tun}) as compared to the $t$-$J$ model are as follows. First, instead of antiferromagnetic nearest-neighbor Heisenberg interactions  ($J_\perp = J_z > 0$), Eq.\ (\ref{eq:tun}) features long-range ($1/R^3$) XXZ interactions with values of $J_\perp$ and $J_z$ that can be independently tuned in magnitude and sign. For example, by adjusting the sign of $J_\perp$, one can obtain the unusual ferromagnetic interactions for fermions and antiferromagnetic interactions for bosons. Second,  instead of the $(-\frac{J}{4} + V) \sum_{\langle i,j\rangle} n_i n_j$ 
interactions, Eq.\ (\ref{eq:tun})  features long-range  ($1/R^3$) density-density interactions ($\propto V$), which can be easily made repulsive to favor 
superfluid correlations. 
Third,  $t$  in Eq.\ (\ref{eq:tun}) can be tuned independently from $J_z$, $J_\perp$, and $V$, and $W$. In particular, one can access the regime $|J_z|, |J_\perp| > t$, which is not possible if $J \sim t^2/U$.  Finally, the term $\propto W$, which describes density-spin interactions, can be made nonzero and can compete with spin-spin interactions ($\propto J_\perp, J_z$) and thus favor new types of spin ordering.


Parameter regimes close to the original $t$-$J$ and $t$-$J$-$V$-$V'$ models can be achieved. In particular, to obtain the model most similar to the $t$-$J$ model, we show in Sec.\ \ref{sec:deriv2} how to set $W = 0$ and $J_z = J_\perp = - 4 V > 0$. We also show how to set $W = 0$, $J_z = J_\perp > 0$, and $V = 0.1 J_z$, which is expected to result in 
a suppression of phase separation relative to the $t$-$J$ model  \cite{troyer93}.  

Being a generalization of the already highly nontrivial $t$-$J$ model (particularly beyond 1D), the Hamiltonian in Eq.\ (\ref{eq:tun}) is expected to give rise to very rich many-body physics. Only a narrow range of this physics has been studied so far. In particular, the Hamiltonians considered in Refs.\ \cite{dallatorre06,burnell09,dalmonte11,goral02,yi07, menotti07,baranov08,burnell09,danshita09,lahaye09,capogrosso-sansone10, pollet10,mikelsons11,he11}
are reminiscent of the restriction of Eq.\ (\ref{eq:tun}) to a single rotational state. The use of more than one rotational state for manybody physics with diatomic polar molecules has been considered before in Refs.\ 
\cite{barnett06, micheli06, micheli07,buchler07,brennen07, buchler07b, gorshkov08c,watanabe09,wall09,yu09b,cooper09,krems09,wall10,schachenmayer10,perezrios10,trefzger10,herrera10,lin10b,kestner11}.
Finally, in Ref.\ \cite{gorshkov11}, using density matrix renormalization group (DMRG) \cite{white92b,white93,schollwock05}, we studied the 1D phase diagram of  the simplest experimentally realizable regime of Eq.\ (\ref{eq:tun}), where $V=W=J_z = 0$,  $t_{m_0} = t_{m_1} = t$, and the two remaining parameters are molecule density and $J_\perp/t$ . As expected from the above discussion, we indeed found an enhancement of 
superfluid correlations and a suppression of phase separation relative to the usual $t$-$J$ model.



Preparation of the phases can be done, for example, by applying an additional microwave field coupling the two dressed rotor states and performing an adiabatic passage from a state that is easy to prepare to the desired 
ground state by tuning the Rabi frequency and the detuning of the microwave field \cite{schachenmayer10}. This extra microwave field, which gives rise to terms proportional to $\sum_j S^x_j$ and $\sum_j S^z_j$ (i.e.\ effective $\mathbf{\hat{x}}$ and $\mathbf{\hat{z}}$ magnetic fields), can also be thought of as a way of enabling the simulation of a richer class of models where $\sum_j S^z_j$ is not conserved. We expect that, by analogy with Ref.\ \cite{schachenmayer10}, preparation of the phases of interest can often be done without single-site addressability. 

Molecules in the rovibrational ground state can be detected by converting them back to atoms \cite{ni08}. Furthermore, efforts towards achieving optical cycling in molecules are under way \cite{wang10,shuman10}. There is, thus, hope that powerful tools for the detection of molecular phases can be borrowed \cite{schachenmayer10} from experiments with ultracold atoms. These tools include noise-correlations in the time-of-flight absorption imaging \cite{altman04,folling05,greiner05} and direct in-situ fluorescent imaging \cite{bakr10,sherson10}. In Ref.\ \cite{gorshkov11}, the possibility of probing the phase diagram of Eq.\ (\ref{eq:tun}) with center-of-mass Bloch oscillations \cite{ben-dahan96,wilkinson96} is also discussed.

The model can be extended to more than two dressed rotor states. 
By applying a sufficient number of microwave fields, one can achieve significant tunability of the 
coefficients even in the resulting more complicated models.


\subsection{Effects of the nuclear degrees of freedom \label{sec:nuc}}

Having discussed the physics of Eq.\ (\ref{eq:nuctun}) in the absence of nuclear spins, we turn in this Section to the discussion of the effects of nuclear spin. One of the simplest Hamiltonians involving nuclear degrees of freedom would be realized in the case of one molecule per site:
 \begin{eqnarray}
&&H = A \sum_i  S^z_i  T^z_i  \nonumber \\
&&+ \frac{1}{2}\sum_{i \neq j} V_\textrm{dd}(\mathbf{R}_i-\mathbf{R}_j)  \left[J_z  S^z_i S^z_j + \frac{J_\perp}{2} (S^+_i S^-_j + S^-_i S^+_j)  \right]. \label{eq:nuc}
\end{eqnarray}
As in Sec.\ \ref{sec:magn}, we ignored edge effects 
and dropped terms commuting with $H$. In this case, $T^z_i$ is conserved on each site and becomes a classical variable. It can play the role of a tunable magnetic field at each site or the role of tunable disorder. The parameter $A$ can be tuned to zero and away from zero, thus, decoupling nuclear spins from the rotor degree of freedom and coupling them. This tuning can be achieved, for example, by changing the strength of the DC electric field (see Secs.\ \ref{sec:hf} and \ref{sec:deriv1}).

While Eqs.\ (\ref{eq:nuctun}) and (\ref{eq:nuc}) feature $S^z_i T^z_i$ hyperfine interactions, other interactions between rotor and nuclear degrees of freedom can also be generated. In particular, we show in Sec.\ \ref{sec:deriv1} that a judicious choice of rotor and nuclear states may allow for interactions of the form $A S^z_i T^z_i + A_2 T^z_i + A_3 T^x_i$ or even $S^+_i T^-_i + S^-_i T^+_i$. Moreover, by combining the Hamiltonian in Eq.\ (\ref{eq:nuctun})  or in Eq.\ (\ref{eq:nuc}) with microwave and/or radio-frequency pulses applied at regular short intervals, one can use the Trotter approximation \cite{suzuki76} to effectively modify the hyperfine interaction between $\mathbf{S}_i$ and $\mathbf{T}_i$ from a simple $S^z_i T^z_i$ interaction to \textit{any} desired form. With this generalization, nuclear spin in Eq.\ (\ref{eq:nuc}) is in general no longer a classical variable. The Hamiltonian would then describe two types of spin-1/2 species (each site having one of each): $S$ species exhibiting interactions with neighboring sites and $T$ species not exhibiting such interactions but interacting with $S$ on the same site. Such a model is 
reminiscent of the Kondo lattice setup \cite{doniach77}.  Moreover, the nuclear spin may allow to simulate the ``orbital" degree of freedom, whose interplay with spin (i.e.\ rotational) and charge (i.e.\ density) degrees of freedom may enable simulations of  the exotic behavior of  spin-incoherent Luttinger liquids \cite{feiguin11}, transition metal oxides \cite{tokura00}, and iron pnictide superconductors \cite{norman08}. Finally, by using more than two nuclear spin states, one might be able, by analogy with alkaline-earth atoms \cite{gorshkov10,cazalilla09}, to simulate exotic high-spin physics. 




While most of the discussion in the present manuscript focuses on quantum magnetism, the system also has promising quantum information applications. Having two outer electrons, alkali dimers have similar electronic structure to that of alkaline-earth atoms. 
Thus, one may consider extending some of the alkaline-earth quantum information processing proposals to polar alkali dimers. In particular, one can extend the idea of encoding quantum information in the nuclear spin degrees of freedom from alkaline-earth atoms \cite{childress05,hayes07,reichenbach07,daley08,gorshkov09,reichenbach09,shibata09} to polar molecules  \cite{rabl06,andre06,rabl07,kuznetsova08}. In this context, by analogy with alkaline-earth quantum register proposals \cite{gorshkov09}, information stored in the nuclear spins can be mapped via hyperfine interactions (or via microwave or radiofrequency fields) onto the rotor degree of freedom, which can then, in turn, be used to couple different molecules. By analogy with Ref.\ \cite{gorshkov09}, we expect this system -- particularly if more than two nuclear spin states are involved -- to be useful in generating high-fidelity many-body entangled states, such as cluster states or squeezed states.



\section{Rotational and hyperfine structure \label{sec:hf}}

Having discussed in Sec.\ \ref{sec:ham} the main features of Eq.\ (\ref{eq:nuctun}), we present in Secs.\ \ref{sec:hf}-\ref{sec:stab} the derivation of Eq.\ (\ref{eq:nuctun}) and the ways, in which the coefficients in Eq.\ (\ref{eq:nuctun}) can be controlled. Since we are interested in the effects of nuclear spin on the many-body Hamiltonian, we begin the derivation of Eq.\ (\ref{eq:nuctun}) by studying in this Section the rotational and hyperfine structure of a single molecule in the presence of a DC electric field and zero or more continuous-wave (CW) microwave fields. The example molecules we are considering are ${}^{40}$K${}^{87}$Rb \cite{ni08} and ${}^7$Li${}^{133}$Cs \cite{deiglmayr08}. 

Following Refs.\ \cite{brown03,aldegunde08,aldegunde09,micheli07,ran10}, the single-molecule Hamiltonian in the presence of a DC electric field and a CW microwave field is
\ba
H &=& H_{0} + H_\textrm{mw} + H_\textrm{hf},
\ea
where
\ba
H_{0} &=& B \mathbf{N}^2 - d_0 E, \label{rr}\\
H_\textrm{mw} &=&   - \mathbf{d} \cdot \left(E_\textrm{mw} \mathbf{e}_\textrm{mw} e^{-i \omega_\textrm{mw} t} + \textrm{c.c.}\right),  \nonumber \\
H_\textrm{hf} &=& H_{Q} + H_{IN} + H_\textrm{t} + H_\textrm{sc} \nonumber \\
&=& - e \sum_{i = 1}^2 T^2(\mathbf{\nabla E}_i) \cdot T^2(\mathbf{Q}_i) + \sum_{i = 1}^2 c_i \mathbf{N} \cdot \mathbf{I}_i  \nonumber \\
&&
- c_3 \sqrt{6} T^2(\mathbf{C}) \cdot T^2(\mathbf{I}_1, \mathbf{I}_2) + c_4 \mathbf{I}_1 \cdot \mathbf{I}_2. \label{eq:hf}
\ea

$H_0$ describes the rigid rotor coupled to the DC electric field. $B$ is the rotational constant and $\mathbf{N}$ is the angular momentum operator describing the rotation of the molecule. The molecular quantization axis is chosen to be $\hat{\mathbf{z}}$, which is the direction of  the applied DC electric field (see Fig.\ \ref{fig:angles}). $\mathbf{d}$ is the dipole moment operator, while $d_p = \hat{\mathbf{e}}_p \cdot \mathbf{d} = d C^1_p(\theta,\phi)$, where $d$ is the permanent dipole moment of the molecule, $p = 0, +1, -1$, and the spherical basis vectors are defined as $ \hat{\mathbf{e}}_0 = \mathbf{\hat{z}}$ and $ \hat{\mathbf{e}}_{\pm 1} = \mp (\mathbf{\hat{x}} \pm i \mathbf{\hat{y}})/\sqrt{2}$ \cite{micheli07}. Here $C^k_p(\theta, \phi) = \sqrt\frac{4 \pi}{2 k + 1} Y_{k,p}(\theta,\phi)$, 
where $Y_{k,p}$ are spherical harmonics, and spherical coordinates $(\theta,\phi)$ describe the orientation of the rotor \cite{micheli07}. 

$H_\textrm{mw}$ describes the coupling of the rotor to a microwave field with amplitude $E_\textrm{mw}$, frequency $\omega_\textrm{mw}$, and polarization  $\mathbf{e}_\textrm{mw}$, which we assume to be equal to $\mathbf{e}_{-1}$, $\mathbf{e}_{0}$, or $\mathbf{e}_{1}$, which stand, respectively, for $\sigma^-$, $\pi$, and $\sigma^+$ polarization relative to the applied DC electric field. While $H_\textrm{mw}$ describes the action of a single microwave field, we will consider below the possibility of applying several microwave fields, in which case $H_\textrm{mw}$ would just feature the sum of the corresponding fields.

$H_\textrm{hf}$ is the hyperfine interaction \cite{brown03,aldegunde08}, which is composed of four contributions: electric quadrupole $H_{Q}$,  spin-rotation $H_{IN}$, tensor $H_\textrm{t}$, and scalar $H_\textrm{sc}$. These Hamiltonians couple the nuclear spins $\mathbf{I_1}$ and $\mathbf{I_2}$ of the two nuclei to $\mathbf{N}$ and to each other. The nuclei are numbered as $1 = \textrm{K}$ and $2 = \textrm{Rb}$ for  ${}^{40}$K${}^{87}$Rb and $1 = \textrm{Li}$ and $2 = \textrm{Cs}$ for ${}^7$Li${}^{133}$Cs. 

The forms of $H_\textrm{t}$ and $H_{Q}$ warrant additional clarification. $H_\textrm{t}$ describes direct and indirect anisotropic interaction between the two nuclei and is a scalar product [see Eq.\ (\ref{eq:sprod})] of two second-rank irreducible spherical tensors. The first tensor, $T^2(\mathbf{I}_1, \mathbf{I}_2)$, is the second-rank tensor formed [see Eq.\ (\ref{eq:tprod})] out of $\mathbf{I}_1$ and $\mathbf{I}_2$. The second tensor, $T^2_p(\mathbf{C}) = C^2_p(\theta, \phi)$, characterizes the orientation of the rotor and, hence, the relative position of the two nuclei. 

$H_Q$ describes the interaction between the electric quadrupole moment of each nucleus $i$ and the electric field gradient at nucleus $i$ due to the electrons and the other nucleus. $H_{Q}$ is also a scalar product of two second-rank irreducible spherical tensors. The first tensor is $T^2(\mathbf{Q}_i) = Q_i \frac{\sqrt{6}}{2 I_i (2 I_i - 1)} T^2(\mathbf{I}_i,\mathbf{I}_i)$, where $T^2(\mathbf{I}_i, \mathbf{I}_i)$, is the second-rank tensor formed out of $\mathbf{I}_i$ and where $eQ_i$ is the electric quadrupole moment of nucleus $i$. The second tensor is $T^2(\mathbf{\nabla E}_i) = - \frac{q_i}{2} T^2(\mathbf{C})$, where $q_i$ characterizes the negative of the electric field gradient at nucleus $i$. The values of all relevant molecular parameters for ${}^{40}$K${}^{87}$Rb and ${}^7$Li${}^{133}$Cs are given in Table \ref{KRbLiCs}. All matrix elements are evaluated in Appendix \ref{sec:matel}.

\begin{table}[b]
\begin{tabular}{|c|c|c|}
\hline
& ${}^{40}$K${}^{87}$Rb & ${}^7$Li${}^{133}$Cs \\
\hline \hline
d (Debye) & 0.566 \cite{ni08} & 5.520 \cite{aymar05} \\
\hline
B (GHz) & 1.114 \cite{ospelkaus10} &  5.636 \\
\hline
B/d (kV/cm) & 3.9 &  2.0 \\
\hline
$d^2/(4 \pi \epsilon_0 (0.5 \mu\textrm{m})^3)$ (kHz) & 0.39 &  37  \\
\hline
$I_1$& 4 & 3/2 \\
\hline
$I_2$ & 3/2 & 7/2 \\
\hline
$(e Q q)_1$ (kHz) & 450 \cite{ospelkaus10} & 18.5  \\
\hline
$(e Q q)_2$ (kHz) & -1410 \cite{ospelkaus10} & 188  \\
\hline
$c_1$ (Hz) & -24.1  & 32  \\
\hline
$c_2$ (Hz) & 420.1 & 3014  \\
\hline
$c_3$ (Hz) & -48.2  & 140  \\
\hline
$c_4$ (Hz) & -2030.4  & 1610  \\
\hline
\end{tabular}
\caption{Molecular parameters for ${}^{40}$K${}^{87}$Rb and ${}^7$Li${}^{133}$Cs. $d$ is the permanent dipole moment, $B$ is the rotational constant, and $I$ is the nuclear spin. $(e Q q)$ characterizes $H_Q$, $c_1$ and $c_2$ characterize $H_{IN}$,  $c_3$ characterizes $H_\textrm{t}$, and $c_4$ characterizes $H_\textrm{sc}$. In $I_i$, $(e Q q)_i$, and $c_{i = 1,2}$, the subscript $i = 1$ stands for K in KRb and for Li and LiCs, while the subscript $i = 2$ stands for Rb in KRb and for Cs in LiCs. The values for ${}^{40}$K${}^{87}$Rb and ${}^7$Li${}^{133}$Cs are taken from Refs.\ \cite{aldegunde08} and \cite{ran10}, respectively, unless otherwise indicated. 
\label{KRbLiCs}} 
\end{table}

\begin{figure}[t]
\begin{center}
\includegraphics[width = \columnwidth]{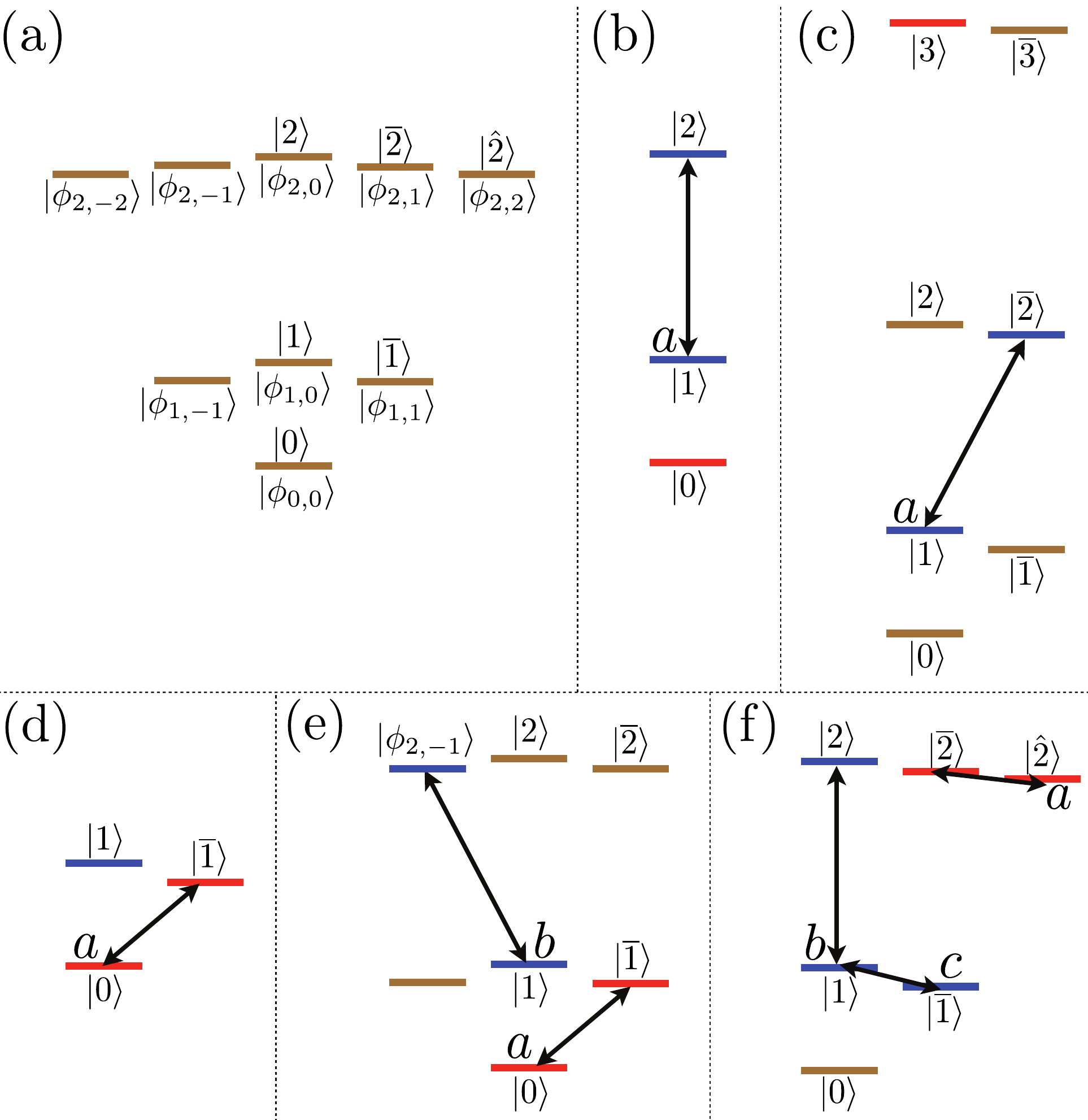}
\caption{(color online). (a) Eigenstates of $H_0 = B \mathbf{N}^2 - d_0 E$. (b-f) Level configurations employing microwaves. The effective two-level dressed rotational degree of freedom $\{|m_0\rangle,|m_1\rangle\}$ is (b) $\{|0\rangle, \sqrt{a} |1\rangle + \sqrt{1-a} |2\rangle\}$, (c) $\{|3\rangle, \sqrt{a} |1\rangle + \sqrt{1-a} |\overline{2}\rangle\}$, (d) $\{\sqrt{a} |0\rangle + \sqrt{1-a} |\overline{1}\rangle,|1\rangle\}$, (e) $\{ \sqrt{a} |0\rangle + \sqrt{1-a} |\overline{1}\rangle, \sqrt{b} |1\rangle + \sqrt{1-b} |\phi_{2,-1}\rangle\}$, (f) $\{\sqrt{a} |\hat{2}\rangle + \sqrt{1-a} |\overline{2}\rangle,\sqrt{b} |1\rangle + \sqrt{c} |\overline{1}\rangle + \sqrt{1-b-c} |2\rangle\}$. In figures (b-f), red (blue) levels make up the effective dressed rotor level $|m_0\rangle$ ($|m_1\rangle$). 
\label{fig:levels}}
\end{center}
\end{figure}

At $E = 0$, the eigenstates of $H_0$ are $|N,M\rangle$ obeying $\mathbf{N}^2 |N,M\rangle = N (N+1) |N,M\rangle$ and $N_z |N,M\rangle = M |N,M\rangle$. As we increase $E$, states with the same $M$ mix to form the new eigenstates. Let us refer to the eigenstate that adiabatically connects to $|N,M\rangle$ (as we turn on $E$) as $|\phi_{N,M}\rangle$, as shown in Fig.\ \ref{fig:levels}(a). While $|\phi_{N,M}\rangle$ are eigenstates of $N_z$ with eigenvalue $M$, they are not eigenstates of $\mathbf{N}^2$ (for nonzero $E$); instead, they are superpositions of $|N',M\rangle$ for different $N'$. To allow for a less cumbersome notation, let us also make the following simplifying definitions illustrated in Fig.\ \ref{fig:levels}(a): $|N\rangle \equiv |\phi_{N,0}\rangle$ and $|\overline{N}\rangle \equiv |\phi_{N,1}\rangle$. In Sec.\ \ref{sec:deriv2}, we will also make use of the definition $|\hat{N}\rangle \equiv |\phi_{N,2}\rangle$ [see Fig.\ \ref{fig:levels}(a)]. The energies of $|\phi_{N,M}\rangle$ and the coefficients in the expansion of $|\phi_{N,M}\rangle$ in terms of $|N',M\rangle$ up to any desired $N$  can easily be computed numerically by truncating the Hilbert space at some other -- much larger -- $N$. The fact that the splitting between $|N,M\rangle$ and $|N+1,M\rangle$ increases with $N$ ensures that for any finite $E$, there will be some $N$ above which the effect of $E$ is negligible. 

\subsection{Hyperfine structure in the simplest level configuration: $\{|m_0\rangle,|m_1\rangle\} = \{|0\rangle,|1\rangle\}$ \label{sec:hf1}}

There is a great variety of possibilities -- especially when microwave fields are applied -- for choosing the two rotational states to play the role of $|m_0\rangle$ and $|m_1\rangle$ in Eq.\ (\ref{eq:nuctun}). In order to make the explanation of the main features of hyperfine structure clearer, we focus in this Section on the simplest example where no microwave fields are applied and where $|m_0\rangle$ and $|m_1\rangle$ correspond to the lowest two $M = 0$ states: $|m_0\rangle = |0\rangle$ ($= |\phi_{0,0}\rangle$) and $|m_1\rangle = |1\rangle$ ($= |\phi_{1,0}\rangle$) [see Fig.\ \ref{fig:levels}(a)]. In Sec.\ \ref{sec:hf2}, we will extend this discussion to other level configurations.

To simplify our effective Hamiltonian, we would like to prevent $H_\textrm{hf}$ from coupling the states $|0\rangle$ and $|1\rangle$ to other states. Therefore,  we need to assume that the applied DC field $E$ is sufficiently large to split $|1\rangle$ from $|\overline{1}\rangle$ and $|\phi_{1,-1}\rangle$  by an amount larger than $H_\textrm{hf}$. E.g., in KRb, to split off $|1\rangle$ 
 from $|\overline{1}\rangle$ and $|\phi_{1,-1}\rangle$ by $|(eQq)_2|$, one needs $dE/B \approx 0.1$. 
Since for KRb, $B/d = 4$ kV/cm, these values of $dE/B$ are readily achievable. For LiCs, the required value of $dE/B$ is even lower [$dE/B = 0.015$] 
since, for LiCs, $|(eQq)_\textrm{2}/B|$ is 40 times smaller. Moreover, in LiCs, $B/d$ is 2 times smaller, which further reduces the required value of $E$. 

Under these assumptions, we can simply project $H_\textrm{hf}$ on the two states $|0\rangle$ and $|1\rangle$, without worrying about the crossterms:
\begin{equation}
H_\textrm{hf} \approx \sum_{m = 0,1} |m\rangle \langle m| \langle m| H_\textrm{hf}|m \rangle. \label{eq:hfproj}
\end{equation}

To understand the consequences of Eq.\ (\ref{eq:hfproj}), let us follow the procedure similar to that in Ref.\ \cite{ran10} and discuss what happens to different terms in $H_\textrm{hf}$ when we take the expectation value in a given rigid rotor state. $\langle m| H_{IN}|m\rangle = 0$ for both states since these are both $M = 0$ states and, therefore, give $\langle m|\mathbf{N}|m\rangle = 0$. $\langle m| H_\textrm{sc}|m \rangle = H_\textrm{sc}$ is unchanged since it does not  involve rigid rotor coordinates. Considering $H_Q$ and $H_\textrm{t}$, we have [using Eqs.\ (\ref{eq:hqfull},\ref{eq:htfull})]
\begin{eqnarray}
\langle H_Q \rangle 
&=& \langle P_2(\cos \theta) \rangle  \sum_{i=1}^2 (e q Q)_i \frac{3 (I^z_i)^2 - I_i (I_i + 1)}{4 I_i (2 I_i - 1)}, \label{eq:Q}\\
\langle H_\textrm{t} \rangle 
&=& c_3 \langle P_2(\cos \theta) \rangle \left( \frac{1}{2} (I_1^+ I_2^- + I_1^- I_2^+) - 2 I_1^z I_2^z \right).
\end{eqnarray}
Here $P_2(\cos \theta) = C^2_0(\theta,\phi)$ is the 2nd degree Legendre polynomial. $T_0^2(Q_i)$ acts on the $i$'th nucleus. We have used Eq.\ (\ref{eq:tprod}) to get explicit expressions for $T_0^2(\mathbf{I}_i,\mathbf{I}_i)$ and $T_0^2(\mathbf{I}_1,\mathbf{I}_2)$. These expressions can also be obtained from Eqs.\ (\ref{eq:tp2ii},\ref{eq:tp2i1i2}).

\begin{figure}[b]
\begin{center}
\includegraphics[width = 0.8 \columnwidth]{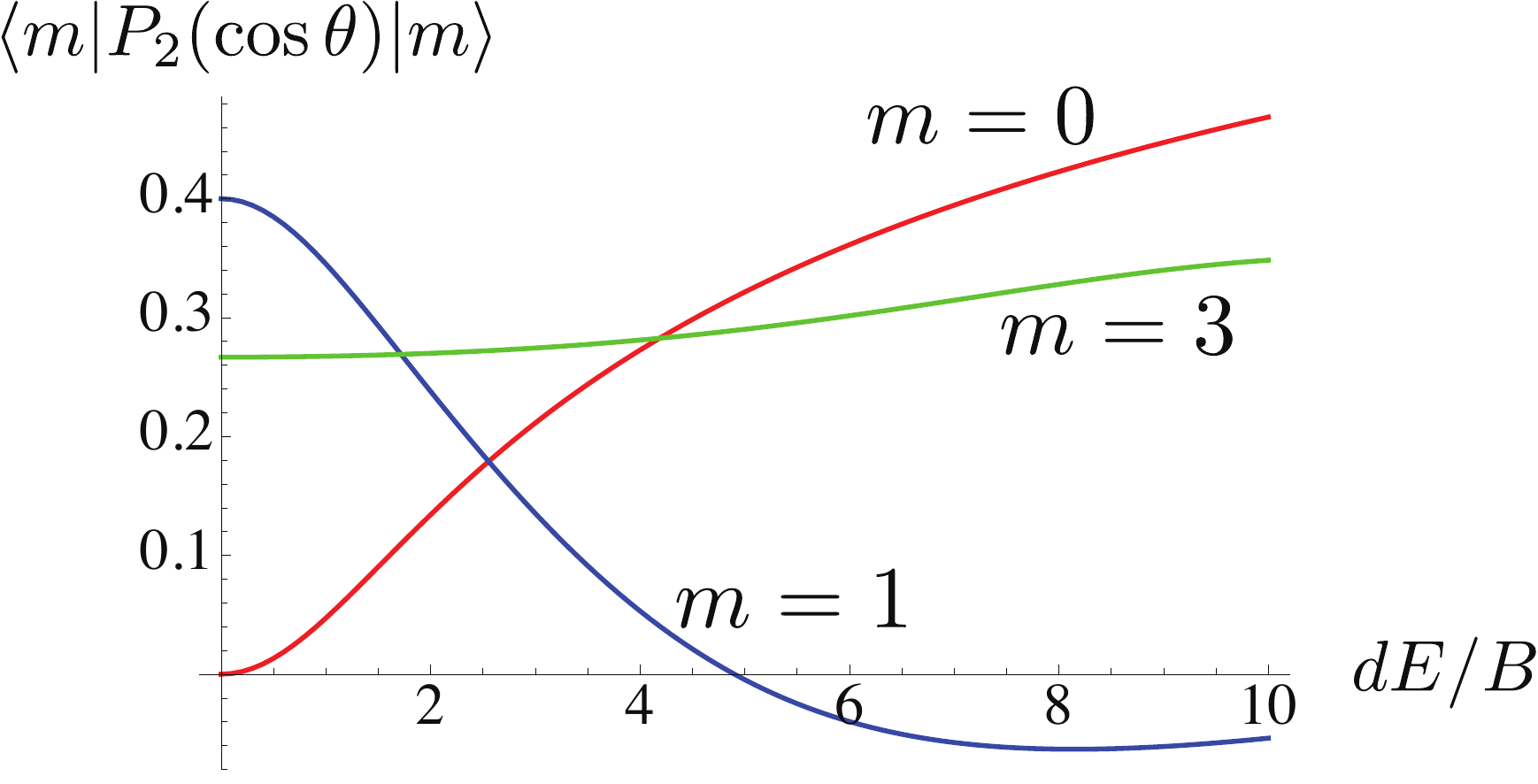}
\caption{(color online). $\langle m| P_2(\cos \theta)|m\rangle = \langle m| C^2_0(\cos \theta)|m\rangle$ as a function of $d E/B$ for $m = 0$ (red), $m = 1$ (blue), and $m = 3$ (green). 
\label{fig:P2}}
\end{center}
\end{figure}

Following Refs.\ \cite{aldegunde08,wall10}, we define the uncoupled basis, in which the two nuclear spin angular momenta are not coupled, and the coupled basis, in which they are coupled to form $\mathbf{I} = \mathbf{I_1} + \mathbf{I_2}$. The matrix elements are evaluated in Appendix \ref{sec:matel} in both bases. We notice that $H_\textrm{sc}$ and $\langle H_\textrm{t} \rangle$ are diagonal in the coupled basis, while $\langle H_Q \rangle$ is diagonal in the uncoupled basis. In Fig.\ \ref{fig:P2},
we plot $\langle m| P_2(\cos \theta)|m\rangle$ for $m = 0, 1, 3$ 
as a function of $dE/B$.  An interesting ``magic" point occurs at $d E/B = 2.55$: $\langle 0|P_2(\cos \theta)|0\rangle = \langle 1|P_2(\cos \theta)|1\rangle = 0.18$, i.e.\ the hyperfine structure in $|0\rangle$ and $|1\rangle$ is exactly the same. $d E/B = 2.55$ means 10 kV/cm for KRb and 5 kV/cm for LiCs, so this point is not easy to access, but it could be useful for both quantum simulation and quantum computation applications as the point of decoupling of the nuclear and rotational degrees of freedom. As we can see in Fig.\ \ref{fig:P2}, similar ``magic" points occur for the pairs of states $\{|0\rangle,|3\rangle\}$ and $\{|1\rangle,|3\rangle\}$ at $d E/B < 10$. 

\begin{figure}[b]
\begin{center}
\includegraphics[width = 0.8 \columnwidth]{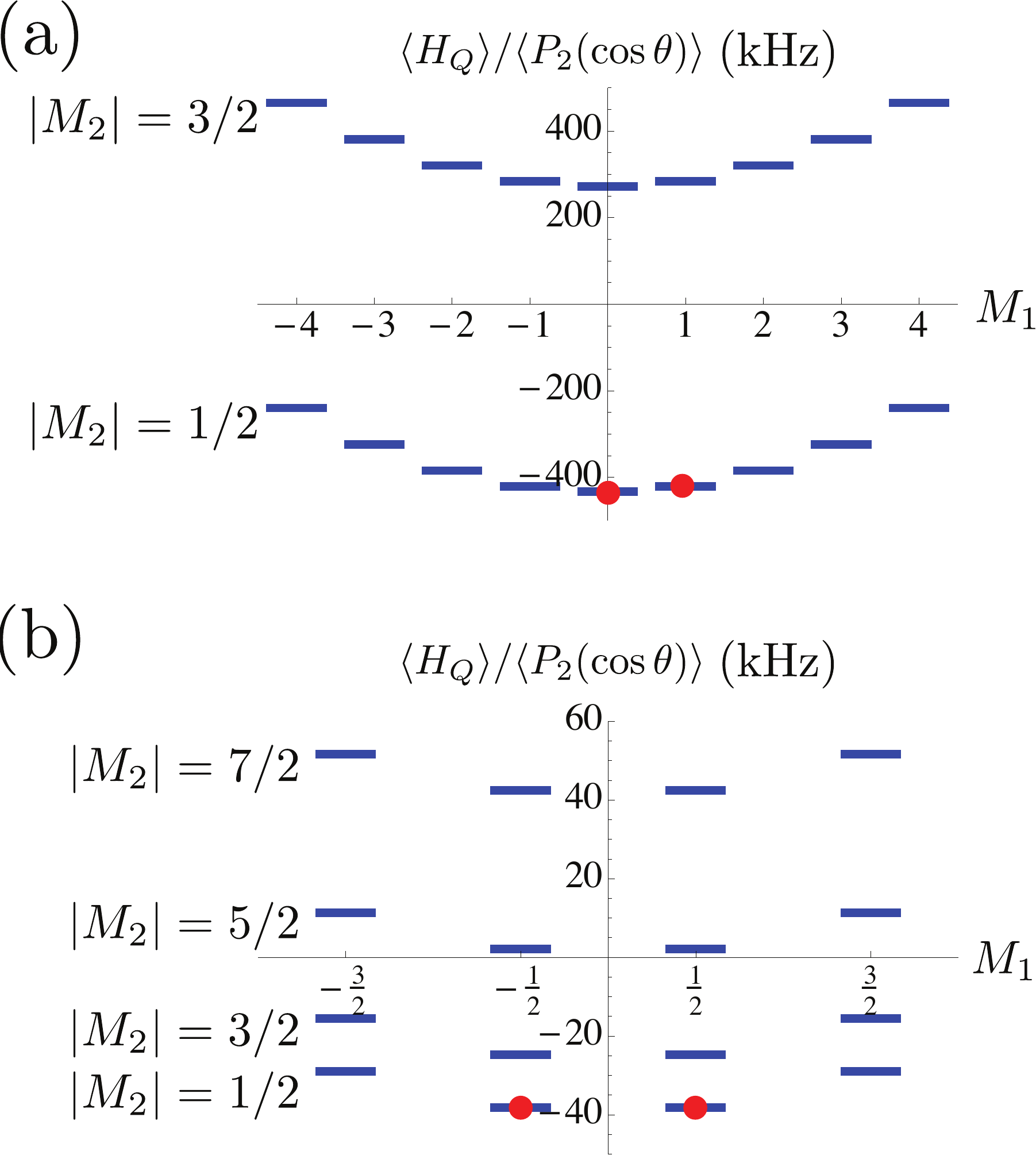}
\caption{(color online). Eigenvalues (in kHz) of $\langle H_Q \rangle/\langle P_2(\cos \theta)\rangle$ for KRb (a) and LiCs (b). The horizontal axis is the magnetic quantum number $M_1$ (for the $K$ nucleus in KRb and for the Li nucleus in LiCs), while $|M_2|$ is indicated separately for each group of levels. The two red circles indicate $(M_1,|M_2|) = (0,1/2)$ and $(1,1/2)$ in (a) and $(M_1,|M_2|) = (-1/2,1/2)$ and $(1/2,1/2)$ in (b). 
\label{fig:hf}}
\end{center}
\end{figure}

Even if we are not at a ``magic" point, where nuclear spin decouples from two rotor states $|0\rangle$ and $|1\rangle$, the hyperfine structure is still relatively easy to understand. $H_Q$ competes with $H_\textrm{sc}$ to determine whether the uncoupled or the coupled basis is a good basis. Since in  KRb (LiCs), $(eQq)_{2}$ is 3 (2) orders of magnitude larger than $c_4$, $\langle m | H_\textrm{hf}| m \rangle$ is almost diagonal in the uncoupled basis, provided $\langle m| P_2(\cos \theta)|m\rangle > 10^{-3} (10^{-2})$. For example, the only place in the range of $dE/B$ values shown in Fig.\ \ref{fig:P2} where this condition breaks down for $|0\rangle$ ($|1\rangle$) is near $dE/B = 0 (5)$, where $\langle m| P_2(\cos \theta)|m\rangle$ goes through zero. As pointed out in Ref.\ \cite{ran10}, at these points,  $\langle m | H_\textrm{hf}| m \rangle = H_\textrm{sc}$. Focusing for the moment on small values of $dE/B$, we find that in KRb (LiCs) $H_\textrm{sc}$ is dominant over $H_Q$ in $|0\rangle$ for $d E/B < 0.1 (0.5)$. On the other hand, we found above that we need $d E/B > 0.1 (0.015)$ to split $|1\rangle$ away from $|\overline{1}\rangle$ by an amount larger than $H_Q$. Thus, near $dE/B = 0$, there is a narrow range of $dE/B$ for LiCs and no such range for KRb, where $|1\rangle$ is sufficiently split from $|\overline{1}\rangle$, but $H_\textrm{sc}$ still dominates the $|0\rangle$ hyperfine structure. This observation supports the statement that for almost all values of $dE/B$, $\langle m|H_\textrm{hf}|m \rangle$ is dominated by $H_Q$, while other hyperfine terms act as a perturbation. Therefore, in Figs.\ \ref{fig:hf}(a,b), 
we show the eigenvalues of $\langle H_Q \rangle/\langle P_2(\cos \theta)\rangle$ for KRb and LiCs, respectively. From Eq.\ (\ref{eq:Q}), we see that these eigenvalues are  $\sum_{i=1}^2 (e q Q)_i \frac{3 (M_i)^2 - I_i (I_i + 1)}{4 I_i (2 I_i - 1)}$, where $M_i$ is the magnetic quantum number of nucleus $i$. 

$H_\textrm{sc}$ and $H_\textrm{t}$ can then be treated as a perturbation:
\ba
&&H_\textrm{sc} + \langle H_\textrm{t} \rangle + \langle H_{IN} \rangle = (c_4 - 2 c_3 \langle P_2(\cos \theta)\rangle) I_1^z I_2^z \nonumber \\ 
&&+ \left(\frac{c_4}{2} + \frac{c_3}{2} \langle P_2(\cos \theta)\rangle\right) (I_1^+ I_2^- + I_1^- I_2^+). \label{eq:pert}
\ea
Here $I_1^z I_2^z$ is diagonal in the uncoupled basis and just shifts the energies slightly. The flip-flop term $I_1^+ I_2^- + I_1^- I_2^+$ changes $(M_1,M_2)$ by $(1,-1)$ or by $(-1,1)$. This term is important provided the energy difference between the two states it connects is smaller than the flip-flop strength. For a typical value of $\langle P_2(\cos \theta)\rangle \sim 0.1$ (see Fig.\ \ref{fig:P2}), the smallest relevant splitting in $\langle H_Q \rangle$ for KRb is between $(M_1,M_2) = (0,1/2)$ and $(1,-1/2)$ [red circles in Fig.\ \ref{fig:hf}(a)] and is equal to about 1 kHz. Since $c_4$ in KRb is $\approx - 2$ kHz, a few of the states in Fig.\ \ref{fig:hf}(a) will get mixed by $H_\textrm{sc} + \langle H_\textrm{t} \rangle$, but for most states the uncoupled basis stays a good basis. The situation is similar in LiCs, where the smallest relevant nonzero splitting in $\langle H_Q \rangle$ is between (1/2,1/2) and (3/2,-1/2). For $\langle P_2(\cos \theta)\rangle \sim 0.1$, this splitting is equal to $\sim 1$ kHz, which is comparable to $c_4 = 1.6$ kHz. An additional feature in LiCs  is  that $\langle H_Q \rangle$ has  degenerate states (-1/2,1/2) and (1/2,-1/2) [red circles in Fig.\ \ref{fig:hf}(b)] that get mixed by the flip-flop term.

It is worth pointing out that the application of a magnetic field can help in defining the uncoupled basis as a good basis, as is done in current experiments \cite{ospelkaus10}. For example, this knob can be used to make the above discussed flip-flops off-resonant or to decouple the nuclear spins when $\langle P_2(\cos \theta)\rangle$ is small. In particular, this allows one to decouple the nuclear spins from each other in state $|0\rangle$ even at small DC electric fields \cite{ospelkaus10}. 

\subsection{Hyperfine structure in other level configurations \label{sec:hf2}}

In the previous Section [Sec.\ \ref{sec:hf1}], we described the hyperfine structure in the simplest configuration of rotational levels: $\{|m_0\rangle,|m_1\rangle\} = \{|0\rangle,|1\rangle\}$. In this Section, we extend this discussion to other configurations of rotational levels.

Even without microwave fields, a great variety of possibilities exist for choosing rotational states to prepare the effective rotor degree of freedom that is featured in Eq.\ (\ref{eq:nuctun}).  One could, for example, choose, instead of states $\{|0\rangle,|1\rangle\}$, the states $\{|1\rangle,|3\rangle\}$, which, as we will see in Sec.\ \ref{sec:stab}, may have some advantages over the former.  

One could also consider $\{|m_0\rangle,|m_1\rangle\} = \{|0\rangle,|\overline{1}\rangle\}$  or $\{|m_0\rangle,|m_1\rangle\} = \{|1\rangle,|\overline{1}\rangle\}$ as the effective rotor degree of freedom. In order to avoid the coupling of $|\overline{1}\rangle$ to $|\phi_{1,-1}\rangle$ by $H_Q$ and $H_t$ (and later by the dipole-dipole interaction), we can apply, for example, a $\sigma^-$-polarized microwave field coupling $|\phi_{1,-1}\rangle$ to $|\phi_{2,-2}\rangle$ that would shift the state $|\phi_{1,-1}\rangle$. Once this is done, $H_\textrm{hf}$ can be projected on each of the two states as in Eq.\ (\ref{eq:hfproj}). Another important difference will be the fact that  $\langle \overline{1}|\mathbf{N}|\overline{1}\rangle = \mathbf{\hat z} \neq 0$, so that $\langle \overline{1}|H_{IN}|\overline{1}\rangle  = \sum_i c_i I^z_i$.  This term will contribute to $\langle \overline{1}|H_\textrm{hf}|\overline{1}\rangle$ in Eq.\ (\ref{eq:pert}). Being diagonal in the uncoupled basis, the term $\langle \overline{1}|H_{IN}|\overline{1}\rangle$ will just slightly shift the levels obtained after diagonalizing $\langle \overline{1}|H_{Q}|\overline{1}\rangle$. This term may provide an extra control knob. In particular, in LiCs, $c_2$ is about twice the value of the scalar coupling $c_4$ and will, thus, play an important role for nuclear spin states that are nearly degenerate under $\langle H_Q \rangle$. In addition to being a control knob, $H_{IN}$ may also give rise to some complications. Specifically, the point where $\langle 1|P_2|1\rangle$ is equal to $\langle \overline{1}|P_2|\overline{1}\rangle$ is, in fact, not an exact magic point for the two states (i.e.\ the two hyperfine structures do not perfectly match) due to the $H_{IN}$ term.  However, first, $c_i$ are rather small (a few orders of magnitude smaller than the dominant quadrupolar term - see Table \ref{KRbLiCs}). Second, $\langle \overline{1}|H_\textrm{IN}|\overline{1}\rangle$ can vanish exactly for $I^z_1 = I^z_2 = 0$, which cannot happen for our isotopes but is, in general, possible. Third, one can slightly adjust the value of the DC electric field from the one that gives $\langle 1|P_2|1\rangle = \langle \overline{1}|P_2|\overline{1}\rangle$ in such a way that some (but not all) desired nuclear spin states have the same relative energies in $|1\rangle$ and $|\overline{1}\rangle$.

The application of microwave fields allows to gain better control over the effective Hamiltonian 
\cite{micheli06, brennen07,buchler07,buchler07b,micheli07,gorshkov08c,yu09b,lin10b,cooper09,wall09,wall10,schachenmayer10,kestner11}. In this Section, we consider two examples of microwave control. In the first example, proposed in Ref.\ \cite{schachenmayer10}, we couple states $|1\rangle$ and $|2\rangle$ with a linearly polarized microwave [Fig.\ \ref{fig:levels}(b)]. We assume that the microwave field is sufficiently weak that it can be treated within the rotating-wave approximation and that its off-resonant couplings on other transitions can be ignored. 
Furthermore, we assume that the microwave Rabi frequency $\Omega = E_\textrm{mw} \langle 2|d_0|1\rangle$ is much larger than the hyperfine structure splittings, so that all hyperfine transitions are addressed equally. In principle, weak microwave fields coupling individual nuclear spin levels can also be used to implement quantum magnetism with polar alkali dimers \cite{micheli06, brennen07, ospelkaus10}; however, for simplicity, we will not discuss this case in the present manuscript.
 In KRb, a Rabi frequency spanning all hyperfine levels (of order a few MHz) requires a microwave intensity of a few W/cm$^2$, which is achievable in the laboratory. In LiCs, which has a 10 times larger dipole moment and 10 times smaller hyperfine splittings, the required microwave intensity is $10^4$ times smaller. The two requirements of staying off-resonant with other rotor transitions (and staying within the rotating-wave approximation) but at the same time addressing all hyperfine levels can easily be achieved since in KRb (LiCs) the splitting between the rotor levels $\sim B$ is 3 (4) orders of magnitudes larger than the largest hyperfine constant $(eQq)_2$ (see Table \ref{KRbLiCs}). 


The application of the microwave field will produce, in the rotating frame, two dressed states \cite{gorshkov08c}. One of them will form the state $|m_1\rangle = \sqrt{a} |1\rangle + \sqrt{1-a} |2\rangle$, where we assumed for simplicity real positive coefficients and 
 where $a$ can be controlled by the amplitude and detuning of the microwave field. Projecting $H_\textrm{hf}$ 
on the subspace spanned by $|m_0\rangle = |0\rangle$ and $|m_1\rangle$, we obtain the following form of the hyperfine interaction
\ba
H_\textrm{hf} &\approx& |0\rangle \langle 0| \langle 0| H_\textrm{hf}|0 \rangle + \nonumber \\
&&+ |m_1\rangle \langle m_1| (a \langle 1| H_\textrm{hf}|1\rangle + (1-a) \langle 2|H_\textrm{hf}|2\rangle).\label{eq:hfproj2}
\ea
Notice that $\langle 1|H_\textrm{hf}|2\rangle$ does not contribute since, in our rotating frame, it is rapidly oscillating. The discussion of the $\{|m_0\rangle,|m_1\rangle\} = \{|0\rangle,|1\rangle\}$ configuration then applies with the change that $\langle 1|H_\textrm{hf}|1\rangle$ is replaced with $a \langle 1| H_\textrm{hf}|1\rangle + (1-a) \langle 2|H_\textrm{hf}|2\rangle$. One advantage of this configuration over the $\{|0\rangle,|1\rangle\}$ configuration is that, for a given choice of the DC electric field,  the magic point where $\langle0|P_2|0\rangle = a \langle 1|P_2|1\rangle + (1-a) \langle2|P_2|2\rangle$ may be accessed by tuning $a$. At this magic point, the nuclear and rotational degrees decouple, as discussed above.

The second example involving microwave fields that we consider in this Section involves the application of a $\sigma^+$ microwave field near resonance with the $|0\rangle-|\overline{1}\rangle$ transition. We pick one of the two rotating-frame dressed states $|m_0\rangle = \sqrt{a} |0\rangle + \sqrt{1-a} |\overline{1}\rangle$ as one of the two effective rotor states and state $|m_1\rangle = |1\rangle$  as the other  [Fig.\ \ref{fig:levels}(d)]. In contrast to the microwave-free $\{|0\rangle,|\overline{1}\rangle\}$ configuration,  in this example, we can safely ignore the state $|\phi_{1,-1}\rangle$ assuming the dressed state $|m_0\rangle$ is shifted by the applied microwave sufficiently far away from the state $|\phi_{1,-1}\rangle$. If $|m_0\rangle$ is too close in energy to $|\phi_{1,-1}\rangle$, then $|\phi_{1,-1}\rangle$ can be shifted away using a separate microwave field coupling it, for example, to $|\phi_{2,-2}\rangle$. Projecting the hyperfine Hamiltonian onto $|m_0\rangle$  and $|1\rangle$, we obtain the same Hamiltonian as in Eq.\ (\ref{eq:hfproj2}) except states $|0\rangle$, $|1\rangle$, and $|2\rangle$ get replaced with states $|1\rangle$, $|0\rangle$, and $|\overline{1}\rangle$, respectively.

Of course, numerous other coupling schemes are also possible. For example, one can apply two microwave fields acting on two different transitions and use one dressed state from each transition as the basis. One can even consider appling more microwave fields, as we will do in Sec.\ \ref{sec:deriv2}. The above discussion of the hyperfine structure can be readily extended to these cases.


\section{Optical potential and tensor shifts \label{sec:opt}}

As discussed in Refs.\ \cite{micheli07,brennen07,wall09,wall10,kotochigova10b}, a rigid rotor placed into an optical lattice experiences level shifts -- called tensor shifts -- that depend on the internal state of the rotor. In this Section, we summarize the derivation of tensor shifts from Ref.\ \cite{micheli07}, consider ways to control these shifts, and discuss the effects of these shifts on our Hamiltonian. 


Following Ref.\ \cite{micheli07}, we consider an off-resonant light field $\mathbf{E}_\textrm{opt}(\mathbf{R},t) = \mathbf{E}_\textrm{opt}(\mathbf{R}) e^{- i \omega t} + \textrm{c.c.}$. We recall that we use the $X$-$Y$-$Z$ coordinates to describe the 2D optical lattice, which lies in the $X$-$Y$ plane, while the $x$-$y$-$z$ coordinates will have $\mathbf{\hat{z}}$ along the applied DC electric field (see Fig.\ \ref{fig:angles}). In the present Section, we will ignore the hyperfine structure -- we will put together the optical potential and the hyperfine structure in Sec.\ \ref{sec:deriv}. The AC Stark shift Hamiltonian acting on a rigid rotor describing the ground electronic and vibrational state of a molecule is then
\ba
H_\textrm{opt}(\mathbf{R}) &=& - \mathbf{E}_\textrm{opt}(\mathbf{R})^* \cdot \hat \alpha(\omega) \cdot \mathbf{E}_\textrm{opt}(\mathbf{R}),
\ea
where
\ba
&&\!\!\!\!\!\!\! \hat \alpha(\omega) 
=   \alpha_{\perp}(\omega) \label{al2}\\ 
 && \!\!\!\!\!\!\! + [ \alpha_{||}(\omega) - \alpha_\perp(\omega)] \sum_{p,p'}(-1)^{p}C^1_{-p}(\theta,\phi) C^1_{p'}(\theta,\phi) \mathbf{\hat{e}}_p \otimes \mathbf{\hat{e}}^*_{p'}.   \nonumber 
\ea
Here $(\theta,\phi)$ are the spherical coordinates of the rotor. 
$\alpha_{||}(\omega)$ and $\alpha_\perp(\omega)$ are dynamical polarizabilities at frequency $\omega$ parallel and perpendicular to the rotor axis. Due to the difference in matrix elements and in the energy difference between states contributing to the two polarizabilities, $\alpha_{||}$ and $\alpha_{\perp}$ are generally different giving rise to the term $\propto  [ \alpha_{||}(\omega) - \alpha_\perp(\omega)]$  describing a rotor-state-dependent shift [second line in Eq.\ (\ref{al2})]. 

We suppose that $\mathbf{E}_\textrm{opt}(\mathbf{R}) = E(\mathbf{R}) \sum^1_{p=-1} \beta_p \mathbf{\hat{e}}_p$, where $\sum_p \beta_p \mathbf{\hat{e}}_p$ is a unit vector (i.e.\ $\sum_p |\beta_p|^2 = 1$) describing the polarization of the light, which, for simplicity, we assume to be spatially uniform. We then find
\ba
H_\textrm{opt}(\mathbf{R}) = - |E(\mathbf{R})|^2 \Big[\frac{2 \alpha_\perp(\omega) + \alpha_{||}(\omega)}{3} \nonumber \\
+ [\alpha_{||}(\omega) - \alpha_\perp(\omega)]\sum_{p=-2}^2 \gamma_p C^2_p(\theta,\phi)\Big],
\ea
where $\gamma_{\pm 2} = -\sqrt{\frac{2}{3}} \beta^*_{\mp 1} \beta_{\pm 1}$, $\gamma_{\pm 1} = \frac{1}{\sqrt{3}} (\beta^*_0 \beta_{\pm 1} - \beta^*_{\mp 1} \beta_0)$, $\gamma_0 = |\beta_0|^2 - \frac{1}{3}$.

We now recall that we will be working at DC electric fields that are large enough to separate the rotor states of interest 
from all the other states by a shift larger than the hyperfine interaction strength ($\gtrsim 1$ MHz). In the cases where the state $|\overline{1}\rangle$ is involved, we assume that a microwave field acting on $|\overline{1}\rangle$ itself or on $|\phi_{1,-1}\rangle$ splits the two by a similarly large shift. Since 1 MHz is greater than typical optical lattice potential strength ($10-100$ kHz), the lattice potential is too weak to induce transitions between the rotor levels, and we can therefore just evaluate $H_\textrm{opt}$ in each rotor state. Moreover, any $|m\rangle = |\phi_{N,M}\rangle$ (with any $M$) is an eigenstate of $N_z$, so, for $p \neq 0$, $\langle m|C^2_p(\theta,\phi)|m\rangle = 0$. Therefore, for such states $|m\rangle$, we get the microwave-free optical potential
\ba
H_\textrm{opt}(\mathbf{R}) = - |E(\mathbf{R})|^2 \Big[\alpha_0(\omega) + \nonumber \\
+ \alpha_2(\omega) \sum_{m} \langle m|P_2(\cos \theta)|m\rangle |m\rangle \langle m|\Big], \label{eq:optproj}
\ea
where
\ba
\alpha_0(\omega) &=& \frac{2 \alpha_\perp(\omega) + \alpha_{||}(\omega)}{3}, \nonumber \\
\alpha_2(\omega) &=&  [\alpha_{||}(\omega) - \alpha_\perp(\omega)] \left(|\beta_0|^2 - \frac{1}{3}\right). \label{eq:al2}
\ea
The dependence of tensor polarizability $\alpha_2(\omega)$ on $\beta_0$ is in direct analogy with the corresponding dependence in atomic tensor polarizabilities \cite{derevianko11,ye08}.

In the case where a microwave field is applied, the optical potential can be computed as follows. For the $\{|m_0\rangle,|m_1\rangle\} = \{\sqrt{a} |0\rangle + \sqrt{1-a} |\overline{1}\rangle,|1\rangle\}$ configuration [Fig.\ \ref{fig:levels}(d)], the optical lattice potential is
\ba
&&H_\textrm{opt}(\mathbf{R}) = - |E_1(\mathbf{R})|^2 \Big[\alpha_0(\omega_1) + \alpha_2(\omega_1) \Big(|1\rangle \langle 1| \langle 1|P_2|1\rangle \nonumber \\
&& + |m_0\rangle \langle m_0| (a \langle 0| P_2|0\rangle + (1-a) \langle \overline{1}|P_2|\overline{1}\rangle)\Big) \Big]. \label{eq:optmw}
\ea
Up to a relabeling of states, a similar expression holds for the $\{|0\rangle, \sqrt{a} |1\rangle + \sqrt{1-a} |2\rangle\}$ configuration [Fig.\ \ref{fig:levels}(b)].

By changing the frequency and the polarization of the applied light and by using several \cite{micheli07,brennen07} laser beams at different frequencies or polarizations, one can control the strength of the tensor shift relative to the scalar shift. In particular, it is often convenient to have a lattice that is independent of the rotor state. One can envision the following avenues for achieving this. 

First, as already pointed out in Ref.\  \cite{kotochigova10b}, for any pair of states $m$ and $m'$, the tensor shift vanishes at the ``magic" points in Fig.\ \ref{fig:P2}, where $\langle m|P_2(\theta)|m\rangle = \langle m'|P_2(\theta)|m'\rangle$ [see Eq.\ (\ref{eq:optproj})]. We recall that these are the same points where the nuclear spins and the rotor degree of freedom decouple. In the case where a microwave field is applied, one has an extra control knob to arrive at the ``magic" point for the two states of interest. For the example considered in Eq.\ (\ref{eq:optmw}), the microwave Rabi frequency and detuning can be used to control  $a$  to obtain a lattice that is the same for states $|m_0\rangle$ and $|1\rangle$, which happens when  $a \langle 0| P_2|0\rangle + (1-a) \langle \overline{1}|P_2|\overline{1}\rangle = \langle 1|P_2|1\rangle$. 

Second, by analogy with ``magic" frequencies for atomic levels \cite{derevianko11,ye08} and for vibrational molecular levels \cite{zelevinsky08}, one may look for a ``magic" frequency $\omega$, for which $\alpha_{||}(\omega) = \alpha_\perp(\omega)$, in which case $\alpha_2(\omega)$ would vanish. However, the search for such a ``magic" frequency may be significantly complicated by the requirement to keep spontaneous emission low \cite{kotochigova06}. 

Third, as already pointed out in Ref.\ \cite{kotochigova10b}, $\alpha_2(\omega)$ would also vanish if one chooses a polarization, such that $|\beta_0|^2 = 1/3$. For example, a linear polarization making an angle $\cos^{-1}(1/\sqrt{3})$ with the $z$-axis (i.e.\ with the DC electric field) would work. 

Fourth, one may use two laser beams \cite{micheli07,brennen07} 
that have $\alpha_2$ of opposite signs. Assuming these beams can be made to have the same spatial profile (which can be done, for example, with holographic techniques \cite{bakr09} or angled beams \cite{nelson07}), their relative intensities can be adjusted in such a way that the combined tensor shift vanishes. 

Finally, if $\alpha_2$ of opposite sign is difficult to achieve, as long as $\alpha_2/\alpha_0$ is different for the two lasers, one can choose the two lasers (on the example of 1D) to have spatial profiles $E_1^2 \cos^2(X K)$ and $E_2^2 \sin^2(X K)$, respectively (for some wavevector $K$). By tuning the relative intensities of the two lasers, one can achieve $E_1^2 \alpha_2(\omega_1) = E_2^2 \alpha_2(\omega_2)$ (where $\omega_1$ and $\omega_2$ are the laser frequencies), which would allow to make the tensor shift spatially independent [$\cos^2 (X K) + \sin^2 (X K) = 1$]. The spatially independent shift can then be treated as a slight modification to the internal structure. 
While this last solution described a 1D lattice, three 1D lattices can be combined into a 3D lattice provided their frequencies differ slightly, so that the lattices do not interfere.

\section{Derivation of the Hamiltonian \label{sec:deriv}}


In this Section, we use the results of Secs.\ \ref{sec:hf} and \ref{sec:opt} to derive the Hamiltonian in Eq.\ (\ref{eq:nuctun}) and to show how various terms in this Hamiltonian can be tuned. We recall that, as shown in Fig.\ \ref{fig:angles}, the molecules are confined to  the $X$-$Y$ plane and are subject to a 2D optical lattice in that plane.  
We also recall that a DC electric field of strength $E$ is applied in the direction $\hat{\mathbf{z}}$ that makes a polar angle $\Theta_0$ with the $Z$-axis and has an azimuthal angle $\Phi_0$ in the $X$-$Y$ plane. The system is then described by five one-body Hamiltonians and one two-body Hamiltonian. The five one-body Hamiltonians are \cite{aldegunde08,aldegunde09,brown03,micheli07,ran10}
\ba
H_0 &=& B \mathbf{N}^2 - d_0 E,\\
H_\textrm{mw} &=&   - \mathbf{d} \cdot \left(E_\textrm{mw} \mathbf{e}_\textrm{mw} e^{-i \omega_\textrm{mw} t} + \textrm{c.c.}\right),\\
H_\textrm{hf}&=&  H_{Q} + H_{IN} + H_\textrm{t} + H_\textrm{sc},\\
H_\textrm{opt} &=& - \mathbf{E}_\textrm{opt}(\mathbf{R})^* \cdot \hat \alpha(\omega) \cdot \mathbf{E}_\textrm{opt}(\mathbf{R}),\\
H_\textrm{kin} &=& \frac{p^2}{2 M_m}.
\ea
The molecules are assumed to be in the electronic and vibrational ground state. $H_\textrm{kin}$ describes the kinetic energy, and $M_m$ is the mass of the molecule (subscript $m$ here stands for the word molecule to avoid confusion with the magnetic quantum numbers $M$).

The two-body Hamiltonian for molecules $1$ and $2$ is given by the dipole-dipole interaction
\ba
H_\textrm{dd} = \frac{1}{4 \pi \epsilon_0 R^3} \left[\mathbf{d}^{(1)} \cdot \mathbf{d}^{(2)} - 3 (\mathbf{\hat R} \cdot \mathbf{d}^{(1)}) (\mathbf{\hat R} \cdot \mathbf{d}^{(2)})\right].
\ea
Here $\mathbf{d}^{(j)}$ is the dipole moment of molecule $j$ and $\mathbf{R} = R \mathbf{\hat R}$ is the vector connecting the two molecules.

From Table \ref{KRbLiCs}, we see that the KRb system has a convenient separation of energy scales, which we have already used in Secs.\ \ref{sec:hf} and \ref{sec:opt}: $H_0 \sim B \sim$ 1 GHz, $H_\textrm{hf} \sim H_{Q} \sim$ 500 kHz, $H_\textrm{opt}+H_\textrm{kin} \sim 10-100$ kHz, $H_\textrm{dd} \sim d^2/(4 \pi \epsilon_0 R^3) \sim 1$ kHz (where we assume a typical separation $R \sim 0.5 \mu$m between two neighboring sites of an optical lattice). As discussed in Sec.\ \ref{sec:hf}, we also choose $H_\textrm{mw}$ to have an energy scale significantly below $H_0$ and significantly above $H_\textrm{hf}$. Therefore, the Hamiltonians can be treated in order of decreasing energy scale. 

In the case of LiCs, dipole-dipole interactions [see Table \ref{KRbLiCs}] are typically on the same order or even stronger than the optical potential. In that case, the physics changes and involves such effects as Wigner crystallization \cite{buchler07,rabl07}. Wigner crytallization has the exciting potential of bringing the molecules closer together (for example, if the optical lattice is not present) and producing strong internal-state-dependent interactions. However, the study of such models involves the phonon modes \cite{rabl07} and is beyond the scope of the present work. 
Therefore, in the present Section, we assume that we either work with KRb or that the rotational levels of LiCs are chosen in such a way [for example, using states $|1\rangle$ and $|3\rangle$ at small DC fields (see Fig.\ \ref{fig:stab}) or employing microwaves] that dipole-dipole interactions are much weaker than the optical potential.


\subsection{Derivation of the Hamiltonian for the simplest level configuration: $\{|m_0\rangle,|m_1\rangle\} = \{|0\rangle,|1\rangle\}$ \label{sec:deriv1}}

To derive the Hamiltonian in Eq.\ (\ref{eq:nuctun}), let us begin in this Section with the simplest case where no microwave fields are applied and where we restrict ourselves to rotor states $|m_0\rangle = |0\rangle$ ($=|\phi_{0,0}\rangle$) and $|m_1\rangle = |1\rangle$ ($=|\phi_{1,0}\rangle$). We will consider other level configurations in Sec.\ \ref{sec:deriv2}.

The diagonalization of $H_0 + H_\textrm{hf} + H_\textrm{mw}$ was discussed in Sec.\ \ref{sec:hf}.  In particular, we showed that for a generic DC electric field, the hyperfine structure in $|0\rangle$ and $|1\rangle$ is almost diagonal in the uncoupled basis. Therefore, if we would like to ignore the nuclear spin, one way to do this is to prepare all molecules in a nuclear spin state that is an eigenstate of both $\langle0| H_\textrm{hf}|0\rangle$ and $\langle 1| H_\textrm{hf}|1\rangle$. The hyperfine energy can slightly change the energy difference between $|0\rangle$ and $|1\rangle$, but the total number of molecules in $|0\rangle$ and the total number of molecules in $|1\rangle$ will be separately conserved, making the precise value of the energy difference between $|0\rangle$ and $|1\rangle$ unimportant. Another way to ignore the nuclear spin is to use the ``magic" point at which the hyperfine structure of $|0\rangle$ and $|1\rangle$ is exactly the same, in which case one does not even have to prepare all molecules in the same nuclear state to observe nuclear-spin-independent dynamics.

To include the nuclear spin into our dynamics in a minimal way, we pick two nuclear spin states $|\uparrow\rangle$ and $|\downarrow\rangle$ that are eigenstates of both  $\langle0|H_\textrm{hf}|0\rangle$  and $\langle1|H_\textrm{hf}|1\rangle$. This can easily be done since, for a generic DC electric field, the two hyperfine structure Hamiltonians are almost diagonal in the same (uncoupled) basis [see Sec.\ \ref{sec:hf}]. In the second quantized notation, we can, thus, write
\begin{equation}
H_0 + H_\textrm{hf} \rightarrow \sum_{m \sigma} E_{m \sigma} n_{m \sigma}, \label{eq:h0hhf}
\end{equation}
where $n_{m \sigma}$ is the number of molecules in internal state $m (= 0,1)$, $\sigma (= \uparrow, \downarrow)$. 

We now consider $H_\textrm{opt} + H_\textrm{kin}$. We suppose that the molecules are confined to the lowest band of a 2D lattice in the $X$-$Y$ plane with the third direction $\mathbf{\hat{Z}}$ frozen out. As discussed in Sec.\ \ref{sec:opt}, $|0\rangle$ and $|1\rangle$ will generically feel lattices of different strength, so that $H_\textrm{opt} = \sum_{m = 0,1} |m\rangle \langle m| V_m(\mathbf{R})$. 
We can then expand the molecular operator $\Psi_{m \sigma}(\mathbf{R})$ in (real) Wannier functions as $\Psi_{m \sigma}(\mathbf{R}) = \sum_j w_{j m}(\mathbf{R}) c_{j m \sigma}$, where $j$ sums over sites in the $X$-$Y$ plane. Here $w_{j m}(\mathbf{R}) = w_{m}(\mathbf{R}-\mathbf{R_j})$, where $\mathbf{R_j}$ is the position of site $j$ in the 2D lattice. Absorbing zero-point energy into Eq.\ (\ref{eq:h0hhf}), $H_\textrm{opt} + H_\textrm{kin}$ can then be rewritten as
\ba
H_\textrm{opt} + H_\textrm{kin} \rightarrow - \sum_{\langle i,j\rangle m \sigma } t_m \left[c^\dagger_{i m \sigma} c_{j m \sigma} + \textrm{h.c.}\right],
\ea
where the sum $\langle i, j \rangle$ is taken over all nearest neighbor pairs and where the tunneling amplitudes are $t_m = - \int d^3 \mathbf{R} w_{i m}(\mathbf{R}) [-  \nabla^2/(2 M_m)  + V_m(\mathbf{R})]  w_{j m}(\mathbf{R})$ for $i$ and $j$ nearest neighbors. For simplicity, we assumed that tunneling amplitudes are the same for all nearest neighbor pairs.

We now consider $H_\textrm{dd}$. Since both $|0\rangle$ and $|1\rangle$ are $M = 0$ states, $H_\textrm{dd}$ (extended to many molecules) simplifies to [see Eqs.\ (\ref{eq:apdd1}-\ref{eq:apdd3})]
\begin{equation}
H_\textrm{dd} = \frac{1}{2} \sum_{i \neq j} V_\textrm{dd}(\mathbf{R_i} - \mathbf{R_j}) d_0^{(i)} d_0^{(j)}. \label{eq:dd}
\end{equation}

In second quantized notation, and keeping only the terms that conserve the total number of molecules in state $m$ (for each $m$), $H_\textrm{dd}$ can be rewritten as
\ba
H_\textrm{dd} &\!\!=\!& \frac{1}{2} \sum_{\sigma \sigma'} \int d^3 \mathbf{R} d^3 \mathbf{R'} V_\textrm{dd}(\mathbf{R}- \mathbf{R'})  \\
&&\times\! \Big\{\!\!\sum_{m m'}\!\! \mu_{m} \mu_{m'} \Psi^\dagger_{m \sigma} (\mathbf{R}) \Psi^\dagger_{m' \sigma'}(\mathbf{R'}) \Psi_{m' \sigma'}(\mathbf{R'}) \Psi_{m \sigma}(\mathbf{R})  \nonumber \\
&&+ \mu_{01}^2 \left[\Psi^\dagger_{0 \sigma} (\mathbf{R}) \Psi^\dagger_{1 \sigma'}(\mathbf{R'}) \Psi_{0 \sigma'}(\mathbf{R'}) \Psi_{1 \sigma}(\mathbf{R}) + \textrm{h.c.}\right]\Big\}, \nonumber
\ea
where $\mu_{m m'} = \langle m|d_0|m'\rangle$ is the transition dipole moment between $|m\rangle$ and $|m'\rangle$ and where $\mu_{m} = \langle m|d_0|m\rangle$ is the dipole moment of state $|m\rangle$. The presence of nonzero dipole moments $\mu_{0}$ and $\mu_{1}$ is expected since $|0\rangle$ and $|1\rangle$ are eigenstates of the rigid rotor Hamiltonian in the presence of a DC electric field. One should keep in mind that certain values of $dE/B$ may give rise to terms that do not conserve the total number of molecules in state $|m\rangle$: for example, at $dE/B \approx 3.24$, the energy difference between $|0\rangle$ and $|1\rangle$ is equal to the energy difference between $|1\rangle$ and $|2\rangle \equiv |\phi_{2,0}\rangle$, and dipole-dipole interactions can resonantly turn two molecules in state $|1\rangle$ into a molecule in state $|0\rangle$ and a molecule in state $|2\rangle$. We assume, however, that we avoid such accidental degeneracies. 

Expanding $\Psi_{m \sigma}(\mathbf{R})$ in Wannier functions, we obtain
\ba
&&\!\!\!\!\!\!\!\!  H_\textrm{dd} = \frac{1}{2} \!\!\!\! \sum_{\textrm{\begin{scriptsize}$\begin{array}{c}j_1 j_2 j_3 j_4 \\ m m' \sigma \sigma'\end{array}$\end{scriptsize}}}  \!\!\!\!  \int d^3 \mathbf{R} d^3 \mathbf{R'} V_\textrm{dd}(\mathbf{R}- \mathbf{R'}) \nonumber \\
&&\!\!\!\!\!\!\!\! \times w_{j_1 m}(\mathbf{R}) w_{j_2 m'}(\mathbf{R'})  w_{j_3 m'} (\mathbf{R'}) w_{j_4 m} (\mathbf{R}) \nonumber \\
&&\!\!\!\!\!\!\!\!  \times    \mu_{m} \mu_{m'}  c^\dagger_{j_1 m \sigma} c^\dagger_{j_2 m' \sigma'} c_{j_3 m' \sigma'} c_{j_4 m \sigma}\nonumber \\
&&\!\!\!\!\!\!\!\!   +  \Bigg[ \frac{1}{2} \!\!\!\! \sum_{\textrm{\begin{scriptsize}$\begin{array}{c}j_1 j_2 j_3 j_4 \\ \sigma \sigma'\end{array}$\end{scriptsize}}} \!\!\!\!  \int d^3 \mathbf{R} d^3 \mathbf{R'} V_\textrm{dd}(\mathbf{R}- \mathbf{R'}) w_{j_1 0}(\mathbf{R}) w_{j_2 1}(\mathbf{R'})  \nonumber \\
&&\!\!\!\!\!\!\!\!  \times w_{j_3 0} (\mathbf{R'}) w_{j_4 1} (\mathbf{R}) \mu_{01}^2 c^\dagger_{j_1 0 \sigma} c^\dagger_{j_2 1 \sigma'} c_{j_3 0 \sigma'} c_{j_4 1 \sigma} + \textrm{h.c.}\Bigg]. \label{eq:hdd0}
\ea
Here the hardcore constraint means that $j_1 \neq j_2$ and $j_3 \neq j_4$.
We now make two approximations: (1) the extent of $w$ is much smaller than the distance between the sites, 
and (2) only terms where $i \equiv j_1 = j_4 \neq j \equiv j_2 = j_3$ contribute. These approximations allow to take $V_\textrm{dd}(\mathbf{R_i}-\mathbf{R_j})$ outside of the integral. 
The result is
\ba
&& \!\!\!\!\!\!\!\!  H_\textrm{dd} = \frac{1}{2} \sum_{i \neq j}V_\textrm{dd}(\mathbf{R_i}- \mathbf{R_j}) \nonumber \\
&& \!\!\!\!\!\!\!\!  \times \left[\sum_{m m'}  \mu_{m} \mu_{m'}  n_{i m} n_{j m'}  + \frac{J_\perp}{2} (S^+_i S^-_j + S^-_i S^+_j) \right], \label{eq:hdd1}
\ea
where $J_\perp = 2 \mu_{01}^2 \left( \int d^3 \mathbf{R} w_{i0}(\mathbf{R}) w_{i1}(\mathbf{R}) \right)^2$. Interestingly, the presence of tensor shifts, thus, does not affect the coefficients of $n_{i m} n_{j m'}$ because the Wannier functions are always normalized. 
The only effect of tensor shifts is, thus, a slight reduction of $J_\perp$ from its tensor-shift-free value of $2 \mu_{01}^2$. The latter makes perfect intuitive sense: the matrix element is reduced due to reduced overlap. One can view this effective modification of $\mu_{01}$ as an extra control knob.  In the remainder of Sec.\ \ref{sec:deriv}, however, we will assume for simplicity that all rotor states feel the same optical potential;  tensor shifts can easily be included by analogy with the above example and will lead to similarly reduced matrix elements.
 Expressing $n_{i m}$ in terms of $n_{i}$ and $S^z_i$, we find that Eq.\ (\ref{eq:hdd1}) is equivalent to $H_\textrm{dd}$ in Eq.\ (\ref{eq:nuctun}) with $V = \frac{(\mu_{0} + \mu_{1})^2}{4}$, $W = \frac{\mu_{0}^2 - \mu_{1}^2}{2}$, and $J_z = (\mu_{0} - \mu_{1})^2$.   In Appendix \ref{sec:iat}, we calculate corrections to the approximations made to arrive at Eq.\ (\ref{eq:hdd1}). While these corrections lead to interesting effects, such as interaction-assisted tunneling, these corrections are small. It is worth pointing out that at  an electric field of $d E/B = 0.1$, which we need to prevent $H_\textrm{hf}$ from coupling $|1\rangle$ to $|\overline{1}\rangle$ and $|\phi_{1,-1}\rangle$ and to decouple the nuclei in state $|0\rangle$, $\mu_{01}^2/\mu_{0}^2 \approx 300$. This means that at this value of $d E/B$, the values of $V$, $W$, and $J_z$ are negligible compared to $J_\perp$, making the $V = W = J_z = 0$ model studied in Ref.\ \cite{gorshkov11} applicable. In row $\#1$ of Table \ref{tab:config}, we collect the values of $V$, $W$, $J_z$, and $J_\perp$ (for the case of no tensor shifts) and list the main features of the $\{|m_0\rangle,|m_1\rangle\} = \{|0\rangle,|1\rangle\}$ scheme.

Let us now simplify the internal state Hamiltonian in Eq.\ (\ref{eq:h0hhf}). Using the definition $n_{i m \sigma} = c^\dagger_{i m \sigma} c_{i m \sigma}$, Eq. (\ref{eq:h0hhf}) can be rewritten as
\begin{equation}
H_0 + H_\textrm{hf}  \rightarrow \sum_{i m \sigma} E_{m \sigma} n_{i m \sigma}. \label{eq:h0hhf2}
\end{equation}
We will use conservation laws to simplify this expression. In particular, our Hamiltonian conserves the total number of $0$ molecules ($n_0 = n_{0 \uparrow} + n_{0 \downarrow}$), the total number of $1$ molecules ($n_1 = n_{1 \uparrow} + n_{1 \downarrow}$), the total number of $\uparrow$ molecules ($n_\uparrow = n_{0 \uparrow} + n_{1 \uparrow}$), and the total number of $\downarrow$ molecules ($n_\downarrow = n_{0 \downarrow} + n_{1 \downarrow}$). Only three out of these four quantities are independent since the first two and the last two quantities both sum to the total number of molecules. Thus, subtracting from the final Hamiltonian constant quantities that commute with it, the only relevant internal-state Hamiltonian will be
\ba
H_{0}+H_\textrm{hf} &\rightarrow& A \sum_{i}  \frac{1}{4} (n_{i 0\uparrow} - n_{i 0 \downarrow} - n_{i 1 \uparrow} + n_{i 1 \downarrow}) \nonumber \\
&=& A \sum_i S^z_i T_i^z, \label{eq:hfA}
\ea
where we assumed that there is at most one molecule per site. Here 
\ba
A &=& (E_{0 \uparrow} - E_{0 \downarrow}) - (E_{1 \uparrow} - E_{1 \downarrow}) \nonumber \\
&\approx& (\langle 0| P_2(\cos \theta)|0\rangle - \langle 1| P_2(\cos \theta)|1\rangle) \nonumber \\
&& \times \sum_{i=1}^2 \frac{3 (e q Q)_i}{4 I_i (2 I_i - 1)} [(M_i)^2 - (M'_i)^2],
\ea
where the last approximation is made provided $H_Q$ dominates the hyperfine structure and where $|\uparrow\rangle = |M_1,M_2\rangle$ and $|\downarrow\rangle = |M'_1, M'_2\rangle$. We see thus that $A$ can be tuned with a significant degree of flexibility. In particular, Fig.\ \ref{fig:P2} shows that $\langle 0| P_2(\cos \theta)|0\rangle - \langle 1| P_2(\cos \theta)|1\rangle$ can be tuned by adjusting $dE/B$. On the other hand, Fig.\ \ref{fig:hf}(a) shows (on the example of KRb) that  
$\sum_{i=1}^2 \frac{3 (e q Q)_i}{4 I_i (2 I_i - 1)} [(M_i)^2 - (M'_i)^2]$
can be adjusted between $12$ kHz [e.g. $(M'_1, M'_2) = (0,1/2)$ and $(M_1, M_2) = (1,1/2)$] and $\sim 1$ MHz.

Let us now briefly discuss the possibility of obtaining more complicated interaction terms between $\mathbf{S}_i$ and $\mathbf{T}_i$ than the simple $A S^z_i T^z_i$ in Eq.\ (\ref{eq:hfA}). First, it is possible to get  a Hamiltonian of the form $A S^z_i T^z_i + A_2 T^z_i + A_3 T^x_i$. 
In the case of one molecule per site in the absence of tunneling, such a Hamiltonian still conserves $S^z_i$ as the original $A S^z_i T^z_i$ Hamiltonian but no longer conserves $T^z_i$. The term $T^x_i$ can be obtained by working in the regime when the term $I_1^+ I_2^- + I_1^- I_2^+$ in Eq.\ (\ref{eq:pert}) couples the two chosen spin states and is not negligible. Whenever $T^x_i$ is not negligible, the term $A_2 T^z_i$ arrises naturally following a derivation similar to that leading to Eq.\ (\ref{eq:hfA}). Second, it is also possible to obtain terms of the form $S^+_i T^-_i + S^-_i T^+_i$. Such terms allow one to exchange $S$ and $T$ excitations within the same molecule. We can obtain such terms  by using, for example, states $|\overline{1}\rangle$ and $|\phi_{1,-1}\rangle$ as the two rotor states. In that case $H_Q$ and $H_\textrm{t}$ can cause transitions between these two levels while at the same time changing $I^z_1 + I^z_2$ by 2.

\begin{table*}[t]
\begin{tabular}{|r|l|l|l|}
\hline
& &\textbf{Expressions for $V$, $W$, $J_z$, $J_\perp$} & \\
&  & $V = [(A_0 + A_1)^2 + B_0 + B_1]/4$ &  \\
&& $W = [A_0^2 + B_0 - A_1^2 - B_1]/2$& \\
& \textbf{Rotor states used} & $J_z = (A_0 - A_1)^2 + B_0 + B_1$ &  \textbf{Special features} \\
\hline
\hline
\#1 &$|m_0\rangle = |0\rangle$ & $V =  (\mu_{0} + \mu_{1})^2/4$ &         -  Simplest.  \\
   &$|m_1\rangle = |1\rangle$ & $W =   (\mu_{0}^2 - \mu_{1}^2)/2$ &   -  At small $dE/B$, $V \approx W \approx J_z \approx 0$ and $J_\perp > 0$,  \\
   &Fig.\ \ref{fig:levels}(a)	& $J_z =(\mu_{0} - \mu_{1})^2$ & $\,\,$ yielding the dipolar $t$-$J_\perp$ Hamiltonain \cite{gorshkov11}.                       \\
   &					& $J_\perp = 2 \mu_{01}^2$ &                 \\
\hline
\#2 &$|m_0\rangle = |1\rangle$ & $V = (\mu_{1} + \mu_{3})^2/4$ &    - At small $dE/B$, $\mu_1 \mu_3 > \mu_{13}^2$,  which may help   \\
   &$|m_1\rangle = |3\rangle$ & $W =   (\mu_{1}^2 - \mu_{3}^2)/2$ & $\,\,$    stabilize the system against chemical reactions                         \\
   &Fig.\ \ref{fig:levels}(a)	& $J_z =(\mu_{1} - \mu_{3})^2$ & $\,\,$ (see Sec.\ \ref{sec:stab}).               \\
   &					& $J_\perp = 2 \mu_{13}^2$ & $\,\,$               \\
\hline 
\#3 &$|m_0\rangle = |0\rangle$ & $V = (\mu_{0} + \mu_{\overline{1}})^2/4$ &  - Simplest configuration with $J_\perp < 0$.       \\
   &$|m_1\rangle = |\overline{1}\rangle$ & $W =   (\mu_{0}^2 - \mu_{\overline{1}}^2)/2$ &  - A microwave field is required to shift $|\phi_{1,-1}\rangle$                           \\
   &Fig.\ \ref{fig:levels}(a)	& $J_z =(\mu_{0} - \mu_{\overline{1}})^2$ & $\,\,$ out of resonance with $|\overline{1}\rangle$.                        \\ 
   &					& $J_\perp =  - \mu_{0\overline{1}}^2$ &                \\
\hline 
\#4 &$|m_0\rangle = |0\rangle$ & $A_0 = \mu_0$    & - At  $(dE/B,a) = (1.25,0.74)$, $W = 0$,      \\
   &$|m_1\rangle = \sqrt{a} |1\rangle + \sqrt{1-a} |2\rangle$ & $A_1 = a \mu_{1} +  (1-a) \mu_{2}$  &  $\,\,$ $J_z = J_\perp  = 0.36 d^2$, and  $V = 0.1 J_z$, making \\
   &Fig.\ \ref{fig:levels}(b)	&  $B_0 = 0$ &  $\,\,$ Eq.\ (\ref{eq:tun})   similar to the SU(2)-symmetric                       \\
   &					&  $B_1 = 2 \mu_{12}^2 a (1-a)$ &  $\,\,$ $t$-$J$-$V$ model, which exhibits suppressed phase             \\
      &					& $J_\perp =  2( \mu_{01}^2 a + \mu_{02}^2 (1-a))$ & $\,\,$ separation \cite{troyer93}.               \\
\hline
\#5 &$|m_0\rangle =\sqrt{a} |0\rangle+\sqrt{1-a}|\overline{1}\rangle$ & $A_0 = a \mu_{0} + (1-a) \mu_{\overline{1}}$  &   - $J_\perp = 0$ can be achieved at any $d E/B$ by \\ 
   &$|m_1\rangle = |1\rangle$ & $A_1 = \mu_{1}$  &  $\,\,$  adjusting $a$. \\ 
   &Fig.\ \ref{fig:levels}(d)	& $B_0 = - \mu_{0\overline{1}}^2 a (1-a)$   & - $J_z < 0$ can be achieved. \\ 
      &					&  $B_1 = 0$ &  - $V < 0$ can be achieved.               \\
   &					& $J_\perp =  2 a \mu_{01}^2  -  (1-a)  \mu_{1\overline{1}}^2$ &                 \\
\hline 
\#6 &$|m_0\rangle =|3\rangle$ & $A_0 = \mu_3$  &   - $J_z = 0$ and $J_\perp = 0$ lines intersect in $(dE/B,a)$      \\
   &$|m_1\rangle =\sqrt{a} |1\rangle + \sqrt{1-a} |\overline{2}\rangle$ &  $A_1 = a \mu_1 + (1-a) \mu_{\overline{2}}$ & $\,\,$ space at $(dE/B,a) = (2.6, 0.92)$. So if                    \\
   &Fig.\ \ref{fig:levels}(c)	& $B_0 = 0$  &  $\,\,$  we write $J_z = |J| \cos \psi$ and $J_\perp = |J| \sin \psi$,                       \\
   &					& $B_1 = -a (1-a) \mu_{1 \overline{2}}$  & $\,\,$  arbitrary $\psi$ can be achieved around that point.                \\
      &					&  $J_\perp = 2 a \mu_{13}^2 - (1-a) \mu_{3 \overline{2}}^2$ &                 \\
\hline 
\#7 &$|m_0\rangle =a |0\rangle +\sqrt{1-a} |\overline{1}\rangle$ & $A_0 = a \mu_0  + (1-a) \mu_{\overline{1}}$  & - At $(dE/B,a,b) = (1.7, 0.33, 0.81)$, $W = 0$ and         \\
   &$|m_1\rangle =b |1\rangle+\sqrt{1-b} |\phi_{2,-1}\rangle$ & $A_1 = b \mu_1  + (1-b) \mu_{\overline{2}}$  & $\,\,$   $J_z = J_\perp = - 4 V = 0.089 d^2$, making Eq.\ (\ref{eq:tun})                        \\
   &Fig.\ \ref{fig:levels}(e)	& $B_0= - a (1-a) \mu_{0\overline{1}}^2 $  &  $\,\,$ very similar to the standard $t$-$J$ model \cite{auerbach94}.                         \\
   &					& $B_1 = - b (1-b) \mu_{1\overline{2}}^2 $   &                 \\
      &					& $J_\perp =  2 a b \mu_{01}^2  - a (1-b) \mu_{0\overline{2}}^2$ &                 \\
      & &   $\quad \quad \,\,\, -  (1-a) b \mu_{\overline{1}1}^2$ & \\
\hline
\#8 &$|m_0\rangle =\sqrt{a} |\hat{2}\rangle + \sqrt{1-a} |\overline{2}\rangle$ & $A_0 = a \mu_{\hat 2}  +(1-a) \mu_{\overline 2}$ &  - The manifolds $V = 0$, $W = 0$, $J_z = 0$,       \\
   & $|m_1\rangle =\sqrt{b} |1\rangle + \sqrt{c} |\overline{1}\rangle $ & $A_1 = b \mu_1  + c \mu_{\overline 1}  + (1-b-c) \mu_2   $ & $\,\,$  and $J_\perp = 0$ intersect in $(dE/B,a,b,c)$ space                           \\
  & $\quad \quad \quad \,\, + \sqrt{1-b-c} |2\rangle $ & $B_0 = - a (1-a) \mu_{{\hat 2}{\overline 2}}^2 $ & $\,\,$  at $(dE/B,a,b,c) = (2.97,0.059,0.56,0.38)$.   \\
   &Fig.\ \ref{fig:levels}(f)	&  $B_1 = - b c \mu_{1{\overline 1}}^2  - c (1-b-c) \mu_{2{\overline 1}}^2$ &  $\,\,$ Full control over $V$, $W$, $J_z$, and $J_\perp$ is                          \\
   &					&  $\quad \quad \,\,\, + 2 b (1-b-c) \mu_{12}^2 $ & $\,\,$ achievable around that point. \\
      &					&  $J_\perp =  2  (1-a) c \mu_{{\overline 2}{\overline 1}}^2  - (1-a) b \mu_{{\overline 2}1}^2$ &                           \\
            &					&  $\quad \quad \,\,\, - (1-a) (1-b-c) \mu_{{\overline 2}2}^2  - a c \mu_{{\hat 2} {\overline 1}}^2$ &                           \\
\hline
\end{tabular}
\caption{The expressions for the dipole-dipole interaction coefficients for several different level configurations. For configurations $\#1$ through $\#3$, the coefficients $V$, $W$, $J_z$, and $J_\perp$ are listed directly. For other configurations, we instead list the expressions for  $A_0$, $A_1$, $B_0$, $B_1$, and $J_\perp$; the expressions for $V$, $W$, and $J_z$ can be computed from $A_p$ and $B_p$ using the formulas provided at the top of the table. While the presented expressions for the interaction coefficients assume no tensor shifts, the effect of tensor shifts is straightforward to include. Some notable features of each configuration are noted in the last column, while a more detailed discussion is provided in the text.   \label{tab:config}}
\end{table*}

\subsection{Derivation of the Hamiltonian for other level configurations \label{sec:deriv2}}

In the previous Section [Sec.\ \ref{sec:deriv1}], we derived the Hamiltonian in Eq.\ (\ref{eq:nuctun}) for the simplest level configuration: $\{|m_0\rangle,|m_1\rangle\} = \{|0\rangle,|1\rangle\}$. In this Section, we show how the coefficients $V$, $W$, $J_z$, and $J_\perp$ in Eq.\ (\ref{eq:nuctun}) can be controlled by choosing other level configurations, including those configurations that involve one or more microwave fields.


The results are summarized in Table \ref{tab:config}. The microwave-free $\{|1\rangle, |3\rangle\}$ scheme ($\#2$ in Table \ref{tab:config}) has the same form of the dipole-dipole coefficients as the $\{|0\rangle,|1\rangle\}$ scheme. However, it has two important features that distinguish it from the $\{|0\rangle,|1\rangle\}$ scheme: the transition dipole moment between $|1\rangle$ and $|3\rangle$ vanishes for $E =0$, and the permanent dipole moments of states $|1\rangle$ and $|3\rangle$ point in the same direction at small fields $E$. This may help stabilize the system against chemical reactions (see Sec.\ \ref{sec:stab}) and may help reduce the strength of dipole-dipole interactions in LiCs below the strength of the optical lattice potential, which is necessary for the applicability to LiCs of the treatment that we present.

To calculate the coefficients $V$, $W$, $J_z$, and $J_\perp$ in the $\{|0\rangle,|\overline{1}\rangle\}$ scheme ($\#3$ in Table \ref{tab:config}), we have to extend Eq.\ (\ref{eq:dd}) to account for the fact that $d_+$ and $d_-$ now play a role (recall that $d_\pm = \mathbf{e}_{\pm1} \cdot \mathbf{d}$). However, energy conservation still forces the conservation of total $N_z$ of the two interacting molecules, thus, making sure that $T_p^2(\mathbf{d}^{(i)}, \mathbf{d}^{(j)})$ contributes only for $p = 0$ [see Eq.\ (\ref{eq:apdd1})]. Therefore, according to Eq.\ (\ref{eq:apdd2}), $d_0^{(i)} d_0^{(j)}$ in Eq.\ (\ref{eq:dd}) should be replaced with $d_0^{(i)} d_0^{(j)} + \frac{1}{2} (d_+^{(i)} d_-^{(j)} + d_-^{(i)} d_+^{(j)})$. Projecting on the states $|0\rangle$ and $|\overline{1}\rangle$ and ignoring off-resonant terms, 
 we obtain
\ba
&& \!\!\!\!\!\!\!\!\!\!\!\!  d_0^{(i)} d_0^{(j)} + \frac{1}{2} \left(d_+^{(i)} d_-^{(j)} + d_-^{(i)} d_+^{(j)}\right) \approx \nonumber \\
&&  \!\!\!\!\!\!\!\!\!\!\!\!   \left(|0\rangle_i \langle 0|_i  \mu_{0} + |\overline{1}\rangle_i \langle \overline{1}|_i \mu_{\overline{1}}\right)  \left(|0\rangle_j \langle 0|_j \mu_{0} + |\overline{1}\rangle_j \langle \overline{1}|_j \mu_{\overline{1}}\right) \nonumber \\
&& \!\!\!\!\!\!\!\!\!\!\!\!   +  \left[- \frac{\mu_{0 \overline{1}}^2}{2} |0 \overline{1}\rangle \langle \overline{1}0| + \textrm{h.c.}\right].\label{eq:dd3}
\ea 
Here $\mu_{m} = \langle m|d_0|m\rangle$, $\mu_{0 \overline{1}} = \langle 0|d_-|\overline{1}\rangle$, and  $|mm'\rangle$ means that molecule $i$ ($j$) is in state $|m\rangle$ ($|m'\rangle$). From now on, we will use the natural notation that $\mu_{m m'} = \langle m|d_p|m'\rangle$ is the transition dipole moment between $|m\rangle$ and $|m'\rangle$ computed using $d_p$ for the appropriate $p$. The interaction in Eq.\ (\ref{eq:dd3}) has the same form as the corresponding interaction for the $\{|0\rangle,|1\rangle\}$ scheme except $\mu_{01}^2$ is replaced with $- \mu_{0\overline{1}}^2/2$. This minus sign comes from the physical effect that two dipoles rotating in the $x$-$y$ plane give an averaged interaction that is equal to negative one-half of the interaction for two dipoles pointing in the $\mathbf{\hat{z}}$ direction \cite{gorshkov08c}. Therefore, this level scheme allows to change the sign of $J_\perp$ relative to the $\{|0\rangle,|1\rangle\}$ scheme. The resulting values for $V$, $W$, $J_z$, and $J_\perp$ are listed in row $\#3$ of Table \ref{tab:config}. 

To calculate the coefficients $V$, $W$, $J_z$, and $J_\perp$ in the $\{|m_0\rangle,|m_1\rangle\} = \{|0\rangle,\sqrt{a} |1\rangle + \sqrt{1-a} |2\rangle\}$ scheme [Fig.\ \ref{fig:levels}(b); $\#4$ in Table \ref{tab:config}], we again ignore off-resonant terms. 
Projecting  $d_0^{(i)} d_0^{(j)}$ onto states $|0\rangle$, $|1\rangle$, and $|2\rangle$, we obtain
\ba
&&\!\!\!\!\!\! d_0^{(i)} d_0^{(j)} \approx (|0\rangle_i \langle 0|_i \mu_{0} + |1\rangle_i \langle 1|_i \mu_{1} + |2\rangle_i \langle 2|_i \mu_{2})  \nonumber \\
&&\!\!\!\!\!\! \times (|0\rangle_j \langle 0|_j \mu_{0} + |1\rangle_j \langle 1|_j \mu_{1} + |2\rangle_j \langle 2|_j \mu_{2}) \nonumber \\
&&\!\!\!\!\!\! + (\mu_{01}^2 |01\rangle \langle 10| + \mu_{02}^2 |02\rangle \langle 20| + \mu_{12}^2 |12\rangle \langle 21| + \textrm{h.c.}).
\ea 
Projecting this on states $|0\rangle$ and $|m_1\rangle$, we arrive at
\ba
&& d_0^{(i)} d_0^{(j)} + \tfrac{1}{2} (d_+^{(i)} d_-^{(j)} + d_-^{(i)} d_+^{(j)}) = \sum_p B_p |m_p m_p\rangle \langle m_p m_p|   \nonumber \\
&&   \sum_{p,q} A_p A_q |m_p m_q\rangle \langle m_p m_q| \!+\! \frac{J_\perp}{2} (|m_0 m_1\rangle \langle m_1 m_0| \!+\! \textrm{h.c.}), \label{eq:dd2}
\ea
where $p,q \in \{0,1\}$ and where the values of $A_p$, $B_p$, and $J_\perp$ are listed in row $\#4$ of Table \ref{tab:config}. Although in the present configuration, $ \tfrac{1}{2} (d_+^{(i)} d_-^{(j)} + d_-^{(i)} d_+^{(j)})$ does not contribute and $B_0 = 0$, we wrote Eq.\ (\ref{eq:dd2}) in this more general form to be able to describe all other configurations below. $A_p$ can be thought of as an effective dipole moment of state $|m_p\rangle$, while $B_p$ gives the contribution to the interaction from the transition dipole moments between the rotor states that make up $|m_p\rangle$. 
Comparing Eq.\ (\ref{eq:dd}) to Eq.\ (\ref{eq:tun}), we can read off $V = [(A_0+A_1)^2 + B_0 + B_1]/4$, $W = (A_0^2 +B_0 - A_1^2  - B_1)/2$, $J_z = (A_0-A_1)^2 + B_0+B_1$. These expressions hold generally and are listed at the top of Table \ref{tab:config}. 
 The two tuning parameters ($dE/B$ and $a$) can be used, for example, to set $W = 0$ and $J_z = J_\perp$. In particular, at $(dE/B,a) = (1.25,0.74)$, we get $W = 0$, $J_z = J_\perp  = 0.36 d^2$ and $V = 0.1 J_z$. Setting $W = 0$ and $J_z = J_\perp$ brings Eq.\ (\ref{eq:tun}) into a form similar to the SU(2)-symmetric $t$-$J$-$V$  model \cite{troyer93} extended to long-range interactions. Moreover, as we have noted in Sec.\ \ref{sec:tjv}, the value of $V = 0.1 J_z$ is expected to result in a suppression of phase separation relative to the original $t$-$J$ model, in which $V = - J_z/4$ \cite{auerbach94}.

To find expressions for $V$, $W$, $J_z$, and $J_\perp$ in the configuration $\{|m_0\rangle,|m_1\rangle\} = \{\sqrt{a} |0\rangle+\sqrt{1-a}|\overline{1}\rangle,|1\rangle\}$ [Fig.\ \ref{fig:levels}(d);  $\#5$ of Table \ref{tab:config}], we project the resonant terms of $d_0^{(i)} d_0^{(j)} + \frac{1}{2} (d_+^{(i)} d_-^{(j)} + d_-^{(i)} d_+^{(j)})$ onto states $|1\rangle$, $|0\rangle$, and $|\overline{1}\rangle$ to obtain
\ba
&& \!\!\!\!\!\!\!\!\!\!\!\!  d_0^{(i)} d_0^{(j)} + \frac{1}{2} (d_+^{(i)} d_-^{(j)} + d_-^{(i)} d_+^{(j)}) \approx \nonumber \\
&&  \!\!\!\!\!\!\!\!\!\!\!\!   (|1\rangle_i \langle 1|_i \mu_{1} + |0\rangle_i \langle 0|_i \mu_{0} + |\overline{1}\rangle_i \langle \overline{1}|_i \mu_{\overline{1}}) \nonumber \\
&& \!\!\!\!\!\!\!\!\!\!\!\!   \times  (|1\rangle_j \langle 1|_j \mu_{1} + |0\rangle_j \langle 0|_j \mu_{0} + |\overline{1}\rangle_j \langle \overline{1}|_j \mu_{\overline{1}}) \nonumber \\
&& \!\!\!\!\!\!\!\!\!\!\!\!   +  (\mu_{01}^2 |01\rangle \langle 10|  - \frac{\mu_{0\overline{1}}^2}{2} |0\overline{1}\rangle \langle \overline{1}0| - \frac{\mu_{1\overline{1}}^2}{2}  |1\overline{1}\rangle \langle \overline{1}1| + \textrm{h.c.}). 
\ea
We note that terms that do not conserve the total $M$ still do not contribute since they are all highly off-resonant (energy non-conserving). In particular, this assumes, that the microwave field is strong enough that $|m_0\rangle$ is not resonant with $|\phi_{1,-1}\rangle$ (or that $|\phi_{1,-1}\rangle$ is shifted away using a separate microwave field coupling it, for example, to $|\phi_{2,-2}\rangle$). Limiting the internal states of the two molecules to $|m_0\rangle$ and $|1\rangle$, we arrive at Eq.\ (\ref{eq:dd2}) with the values of $A_p$, $B_p$, and $J_\perp$ listed in row $\#5$ of Table \ref{tab:config}. 
The minus signs featured in the expressions for $J_\perp$ and $J_z$ (when compared to the $\{|0\rangle, \sqrt{a} |1\rangle + \sqrt{1-a} |2\rangle\}$ configuration - $\#4$ in Table \ref{tab:config}) allow to set $J_\perp = 0$ at any $dE/B$, as well as set $J_z$ and/or $J_\perp$ to be negative. In particular, if one writes $J_z = |J| \cos \psi$ and $J_\perp = |J| \sin \psi$, then the ability to achieve any value of $\psi$ would imply full controllability over $J_z$ and $J_\perp$ and, hence, over Eq.\ (\ref{eq:simple}). And indeed, in a similar configuration $\{|m_0\rangle,|m_1\rangle\}= \{|3\rangle,\sqrt{a} |1\rangle + \sqrt{1-a} |\overline{2}\rangle\}$ [Fig.\ \ref{fig:levels}(c); $\#6$ in Table \ref{tab:config}], by tuning $a$ and $E$, one can achieve any value of $\psi$. In particular, in the plane defined by $d E/B$ and $a$, the $J_z = 0$ and $J_\perp = 0$ lines cross at $(d E/B,a) = (2.6,0.92)$, so that all values of $\psi$ (and hence all four combinations of the signs of $J_z$ and $J_\perp$) can be achieved just by going around that point in a circle. While this proves that any value of $\psi$ can be achieved, the resulting values of $|J|$ could be rather small; however, it is important to emphasize that for any desired $\psi$, there is almost certainly a different level configuration that gives a larger $|J|$.    

To achieve an even larger degree of control, one can apply two microwave fields. For example [Fig.\ \ref{fig:levels}(e); $\#7$ in Table \ref{tab:config}], one microwave field can be used to create a dressed state $|m_0\rangle = a |0\rangle +\sqrt{1-a} |\overline{1}\rangle$, while another microwave field can be used to create a dressed state $|m_1\rangle = b |1\rangle+\sqrt{1-b} |\phi_{2,-1}\rangle$.
Following the same procedure as for other level schemes, we arrive at the expressions for $A_p$, $B_p$ and $J_\perp$ listed in row $\#7$ of Table \ref{tab:config}.
In particular,
with 
$(dE/B,a,b) = (1.7,0.33,0.81)$, we obtain $W = 0$, $J_z = J_\perp = - 4 V = 0.089 d^2$. As we have noted in Sec.\ \ref{sec:tjv}, these values of $W$, $V$, $J_z$, and $J_\perp$ make our model very similar to the original $t$-$J$ model, except the interactions are long-range. Other configurations with two microwave fields can, of course, also be used to obtain other interesting combinations of coefficients or, possibly, to increase the overall interaction strength relative to this example.

Finally, full controllability can be achieved with three microwave fields. In that case, we will have four control knobs (three microwave fields and the magnitude of the applied DC electric field), which can allow for the full control over the four constants $V$, $W$, $J_z$, and $J_\perp$. In particular, consider the example [Fig.\ \ref{fig:levels}(f); $\#8$ in Table \ref{tab:config}] where the two dressed states are $|m_0\rangle = \sqrt{a} |\hat{2}\rangle + \sqrt{1-a} |\overline{2}\rangle$ and $|m_1\rangle = \sqrt{b} |1\rangle + \sqrt{c} |\overline{1}\rangle + \sqrt{1-b-c} |2\rangle$.
To achieve controllability over $a$, we can apply a 
$\sigma^-$ field on the $|\hat{2}\rangle-|\overline{2}\rangle$ transition. 
At $d E/B = 2.97$ (see below), this transition has frequency $0.3 B$ and a sizable transition dipole moment $\mu_{\hat{2} \overline{2}} = -0.13 d$. 
Alternatively, one can use a Raman pair of microwaves to couple $|\hat{2}\rangle$ and $|\overline{2}\rangle$ via $|\hat{3}\rangle \equiv  |\phi_{3,2}\rangle$, in which case the transition dipole moments are stronger ($\mu_{\hat{2} \hat{3}} =  0.37 d$ and $\mu_{\hat{3} \overline{2}} = 0.53 d$) and the transition frequencies are larger ($\sim 6 B$).
To achieve controllability over $b$ and $c$, we can apply a 
 $\sigma^-$ field on the $|\overline{1}\rangle - |1\rangle$ transition (or on the $|\overline{1}\rangle-|2\rangle$ transition) and a $\pi$ field on the $|1\rangle-|2\rangle$ transition. 
In this configuration, we find that the four manifolds $V = 0$, $W = 0$, $J_z = 0$, and $J_\perp = 0$ all intersect at $d E/B = (2.97,0.059,0.56,0.38)$. 
Specifically, to set $V=W=J_z = J_\perp = 0$, it is sufficient to set $A_1 = B_1 = A_0^2 + B_0 = J_\perp = 0$, which is the procedure we followed. Therefore, in a small sphere in the 4-dimensional $(dE/B,a,b,c)$ space  around the intersection point of the four manifolds $V = 0$, $W = 0$, $J_z = 0$, and $J_\perp = 0$, one can achieve any value of $V$, $W$, $J_z$, and $J_\perp$ up to an overall positive prefactor. While this example proves full controllability, the actual magnitude of the interaction could be small in this case; however, it is important to emphasize that for any desired relationship between $V$, $W$, $J_z$, and $J_\perp$, there is almost certainly a different level configuration and a different choice of microwave fields that gives stronger interactions.



The examples presented here (Table \ref{tab:config}) are just a very small fraction of what is possible. In particular, we would like to emphasize that even for the relationships of $V$, $W$, $J_z$, and $J_\perp$ that we consider, configurations other than the ones we present can likely be used to achieve a larger overall interaction strength. Similarly, the search for the optimal configuration for any given experimental laboratory can take into account the laboratory's constraints on the strength of the DC field, on the microwave intensity, and on the range of available microwave frequencies. When designing a configuration to achieve some desired relationship between $V$, $W$, $J_z$, and $J_\perp$, various caveats can be followed to streamline the search. As one simple example of such a caveat, 
one can ensure that some transition dipole moments vanish exactly by using states whose $N_z$ eigenvalues differ by more than one. This way, one can, for example, set $J_\perp = 0$ independently of the strength of the applied DC field. 

\section{Stability against chemical reactions \label{sec:stab}}

We now turn to the discussion of stability of our system against chemical reactions. For some species of diatomic polar molecules, two absolute (electronic, vibrational, rotational, hyperfine) ground state molecules cannot react to form homonuclear dimers \cite{micheli10,zuchowski10}. In that case, one may be able to remove the hard-core constraint and consider Hamiltonians with finite elastic on-site interaction (see e.g.\ Ref.\ \cite{barnett06,he11}). However, even for these molecules, excited states might react  \cite{micheli10}. Moreover, the currently available molecules, KRb and LiCs, both have exothermic reactions to form homonuclear dimers. Therefore, in order to avoid these chemical reactions, it is important to ensure that two molecules never sit on the same site. There are several ways to prevent two molecules from sitting on the same site. First, one can rely on strong dipole-dipole repulsion. Specifically, if we use a 2D geometry in the $X$-$Y$ plane or a 3D geometry with $\mathbf{\hat{Z}}$ tunneling shut off, and if we further suppose that the electric field direction $\mathbf{\hat{z}}$ is near $\mathbf{\hat{Z}}$, 
then, at least for the ground rotational state, 
dipole-dipole repulsion can play the role of a hard-core constraint for molecules when they hop in the $X$-$Y$ plane \cite{buchler07}. One may expect that this repulsion-induced stability also applies to some situations where \textit{two} rotational states are populated. We will discuss this possibility below. Second, sufficiently strong attraction between two molecules that sit on the same site should also be able to prevent two molecules from hopping onto the same site by energy conservation, similar to the experiments on repulsively bound pairs \cite{winkler06}. Finally, if reaction rates \cite{micheli10} are really large, one can also try relying on the quantum Zeno effect to provide the hard-core constraint \cite{syassen08,daley09}. Therefore, if strong attraction and/or the quantum Zeno effect are sufficient to provide stability (i.e.\ strong repulsion is not necessary), our models can be extended to the full 3D geometry with tunneling allowed along all three directions.

\begin{figure}[b]
\begin{center}
\includegraphics[width = 0.8 \columnwidth]{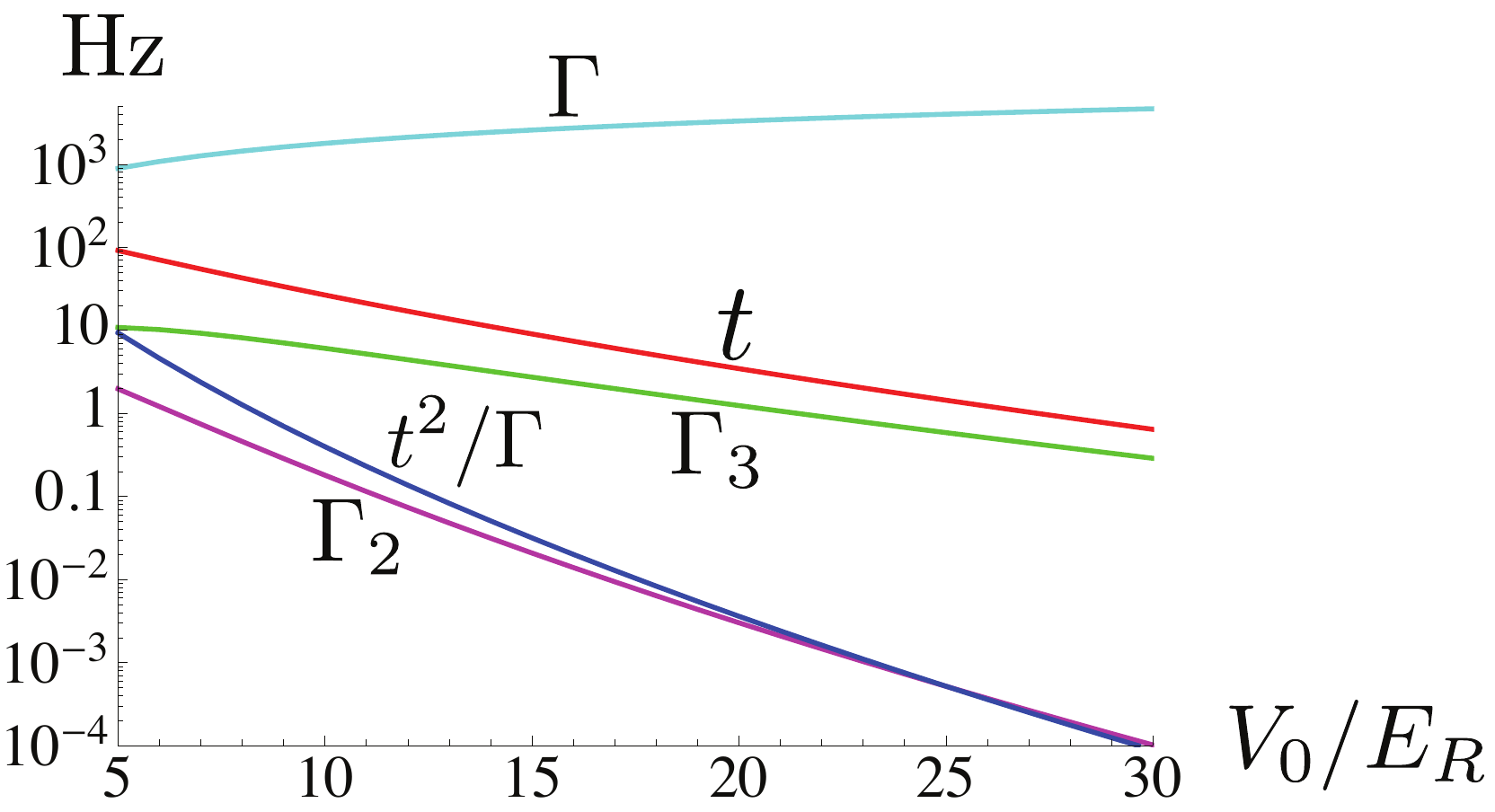}
\caption{(color online). The tunneling amplitude $t$, the on-site chemical reaction rate $\Gamma$, the effective loss rate $t^2/\Gamma$, nearest-neighbor chemical reaction rate $\Gamma_2$, and imaginary part  $\Gamma_3$ of the tunneling amplitude between two occupied sites as a function of $V_0/E_R$, where $E_R$ is the recoil energy and where $V_0$ is the amplitude of the lattice. The vertical axis is in Hz. We use $\lambda = 1064$ nm. \label{fig:decay}}
\end{center}
\end{figure}

Let us make an estimate for the suppression of chemical reactions caused by the quantum Zeno effect.  Let $w(X)$ be the 1D Wannier function for the potential $V_0 \sin^2(K X)$, where $K = 2 \pi/\lambda$ and $\lambda = 1064$ nm \cite{demiranda10}.  We can then compute the tunneling amplitude $t = - \int d X w(X) \left[-\frac{1}{2 m} \frac{d^2}{dX^2}+V_0 \sin^2(K X)\right] w(X-\lambda/2)$ and the on-site chemical reaction rate $\Gamma = \kappa_{3D} \left[\int d X w^4(X)\right]^3$ \cite{garcia-ripoll09}, where we take the 3D loss rate $\kappa_{3D} = 2 \times 10^{-10}$ cm$^{3}$/s from Fig.\ 2B of Ref.\ \cite{ospelkaus10b} (which is of the same order of magnitude as the theoretical predictions of Ref.\ \cite{idziaszek10b}). We plot $t$ and $\Gamma$ in Fig.\ \ref{fig:decay} as a function of $V_0/E_R$, where we used the recoil energy $E_R = \hbar^2 K^2/(2 M_m) \approx (2 \pi) 1.4$ kHz for mass $M_m$ of KRb (recall that $\hbar = 1$). We see that as we increase $V_0/E_R$ from $5$ to $30$, $\Gamma/2 \pi$ grows from $900$ Hz to 5 kHz, while $t/2\pi$ drops from $90$ Hz to 0.6 Hz. Therefore, $t \ll \Gamma$ is satisfied for all the values of $V_0$ considered, and we can compute the effective loss rate   $t^2/\Gamma$ \cite{garcia-ripoll09}, with the result shown in Fig.\ \ref{fig:decay}. We see that even at $V_0 = 5 E_R$, $t^2/\Gamma \approx (2 \pi) 9$ Hz, which is already sufficiently slow to allow for an experiment to be carried out. The effective loss rate falls rapidly to even lower values as we increase $V_0/E_R$ dropping below 1 mHz at $V_0/E_R = 30$. In particular, this means that the simplest $\{|0\rangle,|1\rangle\}$ configuration at small $d E/B$, which gives rise to the $J_z = V = W = 0$ model studied in Ref.\ \cite{gorshkov11}, should be stabilized by the quantum Zeno effect, despite the fact that it is not stabilized by repulsive dipole-dipole interactions (see below).

It is also important to verify that two molecules on neighboring sites would not decay directly due to the overlap of their Wannier functions. To do this, we compute the nearest-neighbor chemical reaction rate $\Gamma_2 = \kappa_{3D} \left[\int d X w^4(X)\right]^2 \int d X w^2(X) w^2(X-\lambda/2)$. As we can see from Fig.\ \ref{fig:decay}, as we increase $V_0/E_R$ from $5$ to $30$, $\Gamma_2/2 \pi$ drops from $2$ Hz to 0.1 mHz, making it negligibly small. We also see that up to $V_0/E_R \approx 30$, $t^2/\Gamma$ is larger than $\Gamma_2$ and, thus, determines the total loss rate. By analogy with the interaction-assisted tunneling discussed in Appendix \ref{sec:iat}, we can also compute the quantity $\Gamma_3 = - \kappa_{3D} \left[\int d X w^4(X)\right]^2 \int d X w^3(X) w(X-\lambda/2)$, which can be thought of as the imaginary part of the tunneling amplitude between two occupied sites. As we can see from Fig.\ \ref{fig:decay}, $\Gamma_3$ is smaller than $t$, and, in particular, much smaller than $\Gamma$. Therefore, we expect the $\Gamma_3$ process to be suppressed in a way similar to the suppression of tunneling $t$ between two occupied sites.

Let us also make a rough estimate for the strength of dipole-dipole interactions for two molecules confined to a single site. Taking the dipole moment $d$ of KRb and a typical distance of $50$ nm between two molecules confined to the same site, we get an interaction energy $E_\textrm{int} = d^2/(4 \pi \epsilon_0 (50\textrm{ nm})^3) \sim (2 \pi) 400$ kHz, which is much larger than the tunneling amplitude $t$ shown in Fig.\ \ref{fig:decay}. In fact, this interaction energy is even larger than the on-site chemical reaction rate $\Gamma$. Therefore, strong dipole-dipole interactions may further suppress the tunneling of molecules between two occupied sites, thus, further reducing the loss due to chemical reactions. However, a more elaborate calculation \cite{micheli10}, which is beyond the scope of this paper, is required to fully understand this effect.


\begin{figure}[b]
\begin{center}
\includegraphics[width = 0.8 \columnwidth]{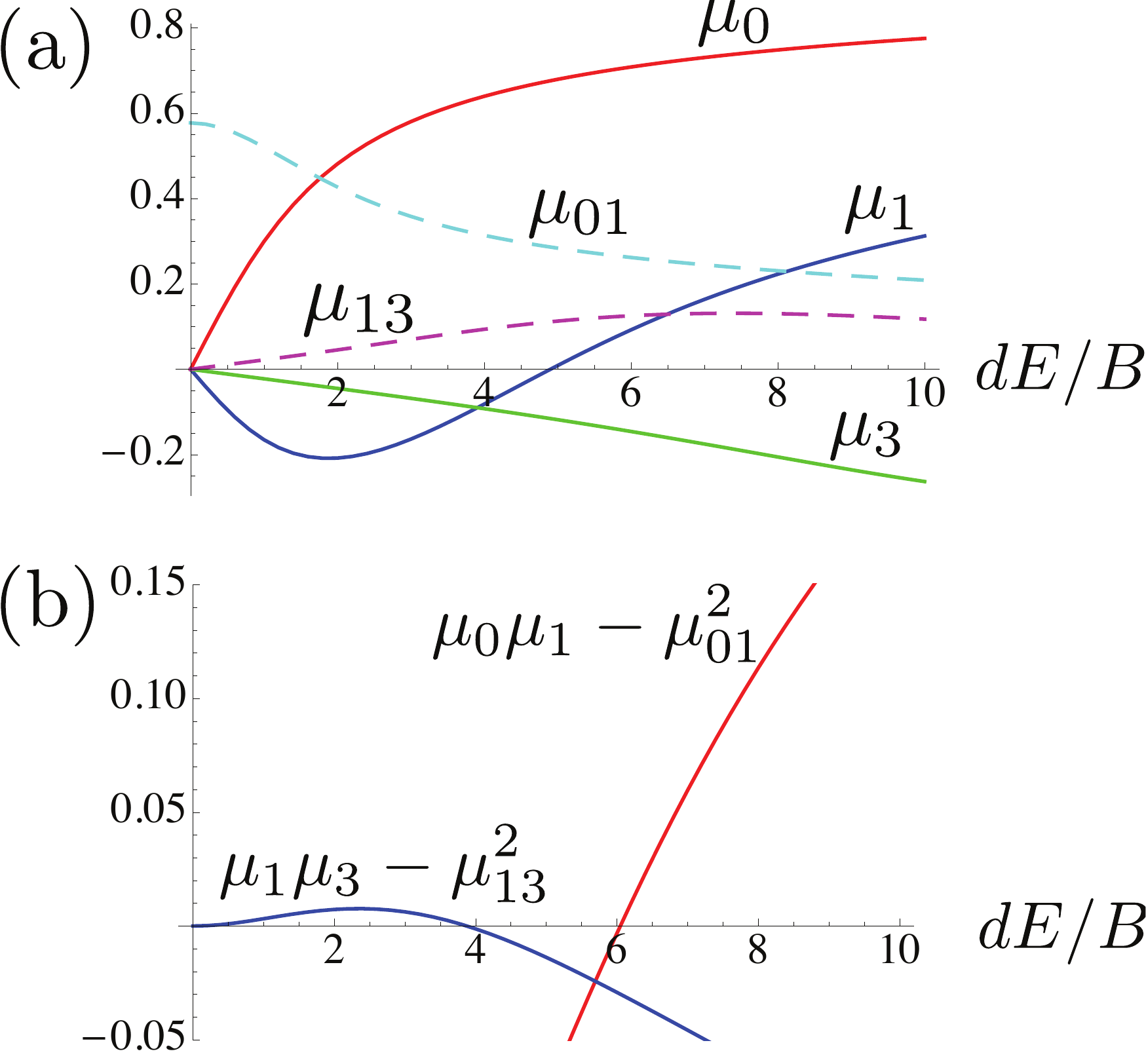}
\caption{(color online). (a) Permanent and transition dipole moments in units of $d$. (b) Stability curves in units of $d^2$. The system is stabilized via repulsive dipole-dipole interactions when the plotted quantity is positive.   \label{fig:stab}}
\end{center}
\end{figure}

Although the quantum Zeno effect and dipole-dipole attraction may allow to stabilize the system (as we have just described), let us, nevertheless, estimate the stability conditions, assuming we want to rely solely on strong \textit{repulsive} dipole-dipole interactions. Since each molecule can be in one of two rotational states, we must ensure repulsion for any two-molecule internal state, which will significantly restrict the range of parameters at which stability is achieved purely by repulsive interactions. Let us begin by considering the simplest $\{|0\rangle,|1\rangle\}$ configuration ($\#1$ in Table \ref{tab:config}). The terms in Eq.\ (\ref{eq:hdd0}) with $j_1 = j_2 = j_3 = j_4$ are the ones that give rise to the hardcore constraint. To ensure dipole-dipole repulsion between two molecules independently of their internal state, two conditions should be satisfied. First, the angle $\Theta_0$ that the DC electric field makes with the $Z$-axis must be smaller than $\sin^{-1}(1/\sqrt{3})$ 
to ensure that $V_\textrm{dd}(\mathbf{R}) > 0$ for any vector $\mathbf{R}$ in the $X$-$Y$ plane [see Fig.\ \ref{fig:angles}]. Second, we have to require that the term in square brackets in Eq.\ (\ref{eq:hdd1}) is positive for any two-molecule internal state.  This requirement reduces to a single condition: $\mu_{0} \mu_{1} > \mu_{01}^2$, where we have assumed $J_\perp =2  \mu_{01}^2$ (i.e.\ no tensor shifts). Physically, this condition ensures that the two-molecule singlet state $(|0\rangle |1\rangle - |1\rangle |0\rangle)/\sqrt{2}$ has positive energy. The same analysis can be done for the $\{|1\rangle,|3\rangle\}$ configuration ($\#2$ in Table \ref{tab:config}) and yields the stability condition $\mu_{1} \mu_{3} > \mu_{13}^2$. In Fig.\ \ref{fig:stab}(a), we show the permanent and transition dipole moments that play a role in these two configurations; and in Fig.\ \ref{fig:stab}(b), we plot the stability curves $\mu_{0} \mu_{1} - \mu_{01}^2$ and $\mu_{1} \mu_{3} - \mu_{13}^2$, whose positive values show the regions of stability. 
We see that, in the $\{|0\rangle,|1\rangle\}$ configuration, stability is achieved for $d E/B > 6$, while, in the $\{|1\rangle,|3\rangle\}$ configuration, it is achieved for $0 < d E/B < 3.9$. The condition $d E/B > 6$  requires large electric fields [$E > 24 (12)$ kV/cm for KRb (LiCs)]. Therefore, it may be easier experimentally to achieve stability in the $\{|1\rangle,|3\rangle\}$ configuration than in the $\{|0\rangle,|1\rangle\}$ configuration. The two features of the $\{|1\rangle,|3\rangle\}$ configuration that allow it to be stable at small DC electric fields are: (1) the fact that $\mu_1$ and $\mu_3$ point in the same direction at small DC fields and (2) the fact that $\mu_{13} = 0$ for $E = 0$. 
We also note that the use of tensor shifts to reduce $J_\perp$ may allow one to extend the stability range to lower $d E/B$ for some configurations, such as the $\{|0\rangle,|1\rangle\}$ configuration.
We also point out that the stability range for the $\{|1\rangle,|3\rangle\}$ configuration conveniently includes the ``magic" point for these two states in Fig.\ \ref{fig:P2} ($d E/B \approx 1.7$). Finally, we note that the analysis in the present Section can be readily extended to the case when microwave fields are applied.




\section{Conclusion \label{sec:conc}}

We derived the $t$-$J$-$V$-$W$ model that governs the behavior of polar alkali dimers in an optical lattice. In particular, we showed how microwave fields can be used to make the coefficients of the Hamiltonian fully tunable. We also described how nuclear spins and the associated hyperfine interactions can be used to endow the model with another highly controllable (orbital) degree of freedom. The peculiar and highly tunable features of the model, such as long-range anisotropic interactions and the hyperfine interactions with the nuclear spin, should make the system an invaluable resource for  studying exotic manybody phenomena and for providing insights into strongly correlated condensed matter systems.

One feature of the models considered in the present manuscript is that, for two nearest-neighbor molecules in an optical lattice with $0.5\mu$m spacing \cite{demiranda10}, dipole-dipole interactions are relatively weak (0.4 kHz for KRb and 37 kHz for LiCs). It would, thus, be convenient to bring the molecules closer.  First, although the structure of molecules is more complicated than that of atoms, and inelastic photon scattering rate could vary 
drastically as one tunes the wavelength of the lattice laser \cite{kotochigova09}, we believe 
that lattice spacing down to 200-300 nm will be possible. This would increase the dipole-dipole interaction strength by an order of magnitude. Another promising way to achieve closely spaced molecules is to consider molecular Wigner crystals \cite{buchler07,rabl07}, which will be the subject of future studies.

Several other extensions of the present work may be particularly fruitful. For example, it is straightforward to extend the Hamiltonian in Eq.\ (\ref{eq:nuctun}) to more than two dressed rotational states and, thus, emulate spin $S > 1/2$. One can also consider level configurations, in which dipole-dipole interaction terms that do not conserve the total $N_z$ of the two interacting molecules contribute [$p \neq 0$ in Eq.\ (\ref{eq:apdd1})] and generate a larger variety of angular dependences in the interaction than that present in $V_{dd}$ [Eq.\ (\ref{eq:vdd})]. Interaction terms of the form $S^+_i S^+_j$ can also be generated if one uses degenerate dressed states $|m_0\rangle$ and $|m_1\rangle$ allowing one to access, for example, spin models beyond the XXZ model. 

Furthermore, while this manuscript is mainly focused on quantum simulation applications of the system, applications to quantum computation -- particularly in the context of storing quantum information in the nuclear spins  \cite{rabl06,andre06,rabl07,kuznetsova08} -- can be readily envisioned.  By analogy with similar proposals for alkaline-earth atoms \cite{gorshkov09}, alkali dimers in an optical lattice may be used, for example, to generate many-body entangled states with applications to precision measurements and to measurement-based quantum computation. Finally, as another possible extension of the present work,  we expect that, by analogy with Ref.\ \cite{schachenmayer10}, which treats Rydberg atoms and polar molecules on equal footing, our ideas should be extendable to Rydberg atoms.

\section{Acknowledgements}

We thank Jun Ye, Peter Zoller, Jesus Aldegunde, Paul Julienne,  Goulven Qu\'em\'ener,  Bernhard Wunsch, Andrew Potter, Alejandro Muramatsu,  Alexander Moreno, Matthias Troyer, Mehrtash Babadi, Peter Rabl, Andrew Daley, and Hendrik Weimer 
for discussions. This work was supported by NSF (Grants No. PHY-0803371, PHY05-51164, PFC, PIF-0904017, DMR-07-05472), the Lee A.\ DuBridge Foundation, ARO with funding from the DARPA OLE program, Harvard-MIT CUA,  AFOSR Quantum Simulation MURI, and AFOSR MURI on Ultracold Molecules.

%



\appendix

\section{Matrix elements \label{sec:matel}}




In this Appendix, we first present some formulas that are useful for evaluating the internal structure of the molecules, their interaction with optical and microwave fields, as well as  their dipole-dipole interaction with each other. We then use these formulas to evaluate matrix elements of the internal molecular Hamiltonian, as well as of the dipole-dipole interaction between two molecules.

Closely following Ref.\ \cite{brown03} for most of this Appendix, let $T^{k_1}_{p_1}(\mathbf{A}_1)$ be a tensor of rank $k_1$ with components $p_1$ which operates on angular momentum $\mathbf{J_1}$. Similarly, let $T^{k_2}_{p_2}(\mathbf{A}_2)$ be a tensor of rank $k_2$ with components $p_2$ which operates on angular momentum $\mathbf{J_2}$. We assume that $\mathbf{J_1}$ and $\mathbf{J_2}$ commute. We can define the tensor product of $T^{k_1}(\mathbf{A}_1)$ and $T^{k_2}(\mathbf{A}_2)$ as
\ba
&&T^{k}_p(\mathbf{A_1},\mathbf{A_2}) = \sum_{p_1} T^{k_1}_{p_1}(\mathbf{A}_1) T^{k_2}_{p - p_1}(\mathbf{A}_2)  (2 k + 1)^{1/2}  \nonumber \\
&& \times  \threej{k_1}{k_2}{k}{p_1}{p-p_1}{-p} (-1)^{-k_1 + k_2 - p}, \label{eq:tprod}
\ea
where the $2\times3$ matrix in parentheses is the $3$-$j$ symbol.
For $k_1 = k_2 = k$, we can also define the scalar product of $T^{k_1}(\mathbf{A}_1)$ and $T^{k_2}(\mathbf{A}_2)$ as 
\ba
T^{k}(\mathbf{A}_1) \cdot T^k(\mathbf{A}_2)  = \sum_p (-1)^p T_p^k(\mathbf{A}_1) T^k_{-p}(\mathbf{A_2}). \label{eq:sprod}
\ea
Notice that  for $k = 1$, the spherical  and Cartesian scalar products agree: $T^{1}(\mathbf{A}_1) \cdot T^1(\mathbf{A}_2) = \mathbf{A}_1 \cdot \mathbf{A}_2$.

If $\mathbf{J_1}$ and $\mathbf{J_2}$ couple to form $\mathbf{J}$, we have the following formulas for the reduced matrix elements (reduced matrix elements use symbol $||$ instead of $|$ and have no dependence on the component indices such as $p_1$, $p_2$, and $p$):
\ba
&&\langle J_1, J_2, J||T^{k_1}(\mathbf{A_1})||J_1',J_2',J'\rangle = \delta_{J_2,J_2'} (-1)^{J' + J_1 + k_1 + J_2} \nonumber \\
&&\times \sqrt{(2 J \!+\! 1) (2 J' \!+\! 1)}\! \sixj{J_1'}{J'}{J_2}{J}{J_1}{k_1} \!\! \langle J_1||T^{k_1}\!(\mathbf{A_1})||J_1'\rangle, \label{eq:tk1}\\
&& \langle J_1, J_2, J||T^{k_2}(\mathbf{A_2})||J_1',J_2',J'\rangle = \delta_{J_1,J_1'} (-1)^{J + J_1 + k_2 + J'_2} \nonumber \\
&& \times \sqrt{(2 J \!+\! 1) (2 J' \!+\! 1)}\! \sixj{J_2'}{J'}{J_1}{J}{J_2}{k_2} \!\! \langle J_2||T^{k_2}\!(\mathbf{A_2})||J_2'\rangle,\label{eq:tk2}\\
&& \langle J_1, J_2, J||T^{k}(\mathbf{A_1},\mathbf{A_2})||J_1',J_2',J'\rangle = \nonumber \\
&& \times \sqrt{(2 J + 1) (2 J' + 1)(2 k + 1)} \ninej{J}{J'}{k}{J_1}{J_1'}{k_1}{J_2}{J_2'}{k_2} \nonumber \\
&&  \times \langle J_1||T^{k_1}\!(\mathbf{A_1})||J_1'\rangle \langle J_2||T^{k_2}\!(\mathbf{A_2})||J_2'\rangle.\label{eq:tk12}
\ea
Here the $2\times3$ matrix in curly braces is the $6$-$j$ symbol, and the $3\times3$ matrix in curly braces is the $9$-$j$ symbol.

The Wigner-Eckart theorem allows to compute matrix elements of $T_p^k(\mathbf{A})$, operating on angular momentum $\mathbf{J}$, in terms of reduced matrix elements: 
\ba
\langle J,M|T_p^k(\mathbf{A})|J',M'\rangle &=& (-1)^{J-M} \threej{J}{k}{J'}{-M}{p}{M'} \nonumber \\
&& \times \langle J||T^k(\mathbf{A})||J'\rangle. \label{eq:we}
\ea

Three particularly useful sets of reduced matrix elements are 
\ba
\langle J||T^1(\mathbf{J})||J'\rangle &=& \delta_{J,J'} [J (J+1)(2 J+1)]^{1/2},\label{eq:jmat}\\
\langle J||T^2(\mathbf{J},\mathbf{J})||J'\rangle &=& \delta_{J,J'}  \frac{J (2 J - 1)}{\sqrt{6}}  \threej{J}{2}{J}{-J}{0}{J}^{-1} \label{eq:jjmat}
\ea
for any angular momentum $\mathbf{J}$ (in particular, for $\mathbf{N}$) and
\ba 
&& \langle N ||T^k(\mathbf{C})||N' \rangle =  (-1)^{N} [(2 N + 1)(2 N' + 1)]^{1/2} \nonumber \\
&&  \times  \threej{N}{k}{N'}{0}{0}{0}, \label{eq:cred}
\ea
where $T^k_p(\mathbf{C}) = C^k_p(\theta,\phi)$.

We will now use Eqs.\ (\ref{eq:tprod}-\ref{eq:cred}) to evaluate matrix elements of the internal molecular Hamiltonian, as well as of the dipole-dipole interaction between two molecules. In Ref.\ \cite{aldegunde08}, which we follow together with Ref.\ \cite{brown03} to compute the matrix elements, three kinds of bases are used. Since we work in the regime where it is sufficient to take the expectation value of $H_\textrm{hf}$ in a given eigenstate of $H_0$, we will use only two basis sets, as in Ref.\ \cite{wall10}: 
\begin{eqnarray}
&&|N M M_1 M_2 \rangle \quad \textrm{(uncoupled)},\\
&&|N M I M_I\rangle \quad \textrm{(coupled)}.
\end{eqnarray}
In both bases, the rotor state $|N M\rangle$ is decoupled from the nuclear spin states $|I_1 M_1\rangle$ and $|I_2 M_2\rangle$. The coupled basis couples the two nuclear spins and uses $|I M_I\rangle$, where $\mathbf{I} = \mathbf{I_1} + \mathbf{I_2}$, while in the uncoupled basis $M_1$ and $M_2$ magnetic quantum numbers of the two nuclear spins are used. We will use the two bases whenever the operator that is being considered acts on the nuclear spins. Otherwise -- if the operator acts only on the rotor degree of freedom -- we will simply use the basis $|N M\rangle$.


We begin by computing the matrix elements of $H_0$ 
[Eq.\ (\ref{rr})]. Noting that $\mathbf{N}^2$ acts only on the rotor degree of freedom, we have
\ba
&&\langle N M|\mathbf{N}^2 |N' M'\rangle =  \delta_{N N'} \delta_{M M'} N (N+1).
\ea
To evaluate the matrix elements of $d_0$, we note that $d_p = T^1_p(\mathbf{d}) = \mathbf{\hat{e}}_p \cdot \mathbf{d} = d C^1_p(\theta,\phi)$ for all 3 values of $p = 0, \pm 1$. Thus, for evaluating the DC Stark shift $-d_0 E$ and the dipole-dipole interaction between two molecules, we need the matrix elements of $C^1_p(\theta,\phi)$. For evaluating the quadrupole hyperfine interaction and the tensor hyperfine interaction, we need the matrix elements of $C^2_p(\theta,\phi)$. Let us, thus, evaluate the matrix elements of $C^k_p(\theta,\phi)$ for general $k$. Using Eqs.\ (\ref{eq:we},\ref{eq:cred}), we have
\ba
&& \langle N M|C^k_p(\theta,\phi)|N' M'\rangle =  (-1)^{M} [(2 N + 1)(2 N' + 1)]^{1/2} \nonumber \\
&&  \times \threej{N}{k}{N'}{-M}{p}{M'} \threej{N}{k}{N'}{0}{0}{0} .
\ea

We now compute the matrix elements of $H_\textrm{hf}$ [Eq.\ (\ref{eq:hf})]. We begin with $H_Q$. Using the form of $H_Q$ in Eq.\ (\ref{eq:hf}), the definitions of $T^2(\mathbf{\nabla E}_i)$ and $T^2(\mathbf{Q}_i)$, and Eq.\ (\ref{eq:sprod}), we have
\ba
H_Q = \sum_{p,i} (-1)^p C_p^2(\theta, \phi) \frac{\sqrt{6} (e q Q)_i}{4 I_i (2 I_i - 1)} T_{-p}^2(\mathbf{I}_i, \mathbf{I}_i), \label{eq:hqfull}
\ea
where $i$ sums over the two nuclei. Since we have already evaluated the matrix elements of $C_p^2(\theta,\phi)$, it remans to list the matrix elements of $T_{p}^2(\mathbf{I}_i, \mathbf{I}_i)$.
Using Eqs.\ (\ref{eq:we},\ref{eq:jjmat}), in the uncoupled basis, they are 
\ba
&&\langle M_i |T_{p}^2(\mathbf{I}_i, \mathbf{I}_i)|M_i'\rangle =  \frac{I_i (2 I_i - 1)}{\sqrt{6}}  \nonumber\\
&&\times (-1)^{I_i - M_i} \threej{I_i}{2}{I_i}{-M_i}{p}{M_i'}  \threej{I_i}{2}{I_i}{-I_i}{0}{I_i}^{-1}. \label{eq:tp2ii}
 \ea
Using Eqs.\ (\ref{eq:we},\ref{eq:tk1},\ref{eq:tk2},\ref{eq:jjmat}), in the coupled basis, they are
\ba
&& \langle I M_I|T_{p}^2(\mathbf{I}_i, \mathbf{I}_i)|I' M_I'\rangle =  \nonumber \\
&& \times   (-1)^{I- M_I + I_1 + I_2} [ (2 I +1) (2 I' + 1)]^{1/2} \frac{I_i (2 I_i - 1)}{\sqrt{6}} \nonumber \\ 
&& \times    \threej{I}{2}{I'}{-M_I}{p}{M_I'}  \threej{I_i}{2}{I_i}{-I_i}{0}{I_i}^{-1} \nonumber \\
&& \times 
            \sixj{I_i}{I'}{I_j}{I}{I_i}{2} 
\Big\{\begin{array}{c}
          (-1)^{I'} \quad \textrm{ if } i = 1\\
          (-1)^{I}  \quad \textrm{ if } i = 2
          \end{array},
\ea
where $j = 2 (1)$ if $i = 1 (2)$. 


To compute the matrix elements of $H_{IN}$, it is sufficient [by Eq.\ (\ref{eq:sprod})] to compute the matrix elements of $\mathbf{N}$ and $\mathbf{I}_i$. Using Eqs.\ (\ref{eq:we},\ref{eq:jmat}), we find that the matrix elements of $\mathbf{N}$ are   
\ba
&& \langle N M|T_p^1(\mathbf{N})|N' M'\rangle = \delta_{N,N'} (-1)^{N - M} \threej{N}{1}{N'}{-M}{p}{M'} \nonumber \\
&& \times [N (N+1) (2 N + 1)]^{1/2}.
\ea
The matrix elements of $\mathbf{I}_i$ in the uncoupled basis are [using Eqs.\ (\ref{eq:we},\ref{eq:jmat})]
\ba
&& \langle M_i|T_p^1(\mathbf{I}_i)|M'_i\rangle = (-1)^{I_i - M_i} \threej{I_i}{1}{I'_i}{-M_i}{p}{M'_i} \nonumber \\
&& \times [I_i (I_i+1) (2 I_i + 1)]^{1/2}. \label{eq:tp1}
\ea
In the coupled basis, they are [using Eqs.\ (\ref{eq:we},\ref{eq:tk1},\ref{eq:tk2},\ref{eq:jmat})]
\ba
&& \langle I M_I|T_p^1(\mathbf{I}_i)|I' M_I'\rangle \!=\! 
- (-1)^{I - M_I+I_1 + I_2} \! \threej{I}{1}{I'}{-M_I}{p}{M_I'} \nonumber \\
&& \times [(2 I +1) (2 I' + 1)]^{1/2}
\! \sixj{I_i}{I'}{I_j}{I}{I_i}{1} \! [I_i(I_i +1)(2 I_i + 1)]^\frac{1}{2}\nonumber \\
&& \times \Big\{\begin{array}{c}
          (-1)^{I'} \quad \textrm{ if } i = 1\\
          (-1)^{I}  \quad \textrm{ if } i = 2
          \end{array},
\ea
where $j = 2 (1)$ if $i = 1 (2)$.


We now turn to $H_\textrm{t}$. Using Eq.\ (\ref{eq:sprod}), 
\ba
H_\textrm{t} = - c_3 \sqrt{6} \sum_p (-1)^p C^2_{-p}(\theta, \phi) T^2_{p}(\mathbf{I}_1,\mathbf{I}_2). \label{eq:htfull}
\ea
Thus, since we have already evaluated the matrix elements of $C^2_p$, it remains to evaluate the matrix elements of $T^2_{p}(\mathbf{I}_1,\mathbf{I}_2)$. In the uncoupled basis, they are [using Eqs.\ (\ref{eq:tprod},\ref{eq:tp1})]
\ba
&& \!\!\!\! \langle M_1 M_2|T^2_{p}(\mathbf{I}_1,\mathbf{I}_2)|M_1' M_2'\rangle =   (-1)^{I_1 -M_1 + I_2 - M_2-p} \nonumber \\
&& \!\!\!\! \times [5 I_1 (I_1 + 1) (2 I_1 + 1)I_2 (I_2 + 1) (2 I_2 + 1)]^{1/2}    \nonumber \\
&& \!\!\!\! \times \sum_{p_1=-1}^1 \threej{1}{1}{2}{p_1}{p-p_1}{-p} \threej{I_1}{1}{I_1}{-M_1}{p_1}{M_1'} \nonumber \\
&& \!\!\!\! \times \threej{I_2}{1}{I_2}{-M_2}{p-p_1}{M_2'}. \label{eq:tp2i1i2}
\ea
In the coupled basis, they are [using Eqs.\ (\ref{eq:we},\ref{eq:tk12},\ref{eq:jmat})] 
\ba
&&\!\!\!\!\!\!\!\! \langle I M_I|T^2_{p}(\mathbf{I}_1,\mathbf{I}_2)| I' M_I'\rangle =  (-1)^{I - M_I}  [5(2 I + 1) (2 I' + 1) ]^\frac{1}{2} \nonumber \\
&&\!\!\!\!\!\!\!\!  \times  [ I_1 (I_1 + 1) (2 I_1 + 1) I_2 (I_2 + 1) (2 I_2 + 1)]^{1/2} \nonumber\\
&&\!\!\!\!\!\!\!\! \times    \threej{I}{2}{I'}{-M_I}{p}{M_I'} \ninej{I}{I'}{2}{I_1}{I_1}{1}{I_2}{I_2}{1}.
\ea


Finally, the matrix elements of $H_\textrm{sc}$ in the uncoupled basis are [using Eqs.\ (\ref{eq:sprod},\ref{eq:tp1})]
\ba
&& \!\!\!\!\!\!\!\! \langle M_1 M_2|H_\textrm{sc}|M_1' M_2'\rangle = c_4 (-1)^{I_1 - M_1 + I_2 - M_2} \nonumber \\
&& \!\!\!\!\!\!\!\! \times   [I_1(I_1 +1)(2 I_1 + 1)I_2(I_2 +1)(2 I_2 + 1)]^\frac{1}{2} \nonumber\\
&& \!\!\!\!\!\!\!\! \times  \sum_{p=-1}^1 (-1)^p \threej{I_1}{1}{I_1}{-M_1}{p}{M_1'} \threej{I_2}{1}{I_2}{-M_2}{-p}{M_2'}.
\ea
In the coupled basis, they are ($|I_1 - I_2| \leq I \leq I_1 + I_2$ is assumed)
\ba
&&\!\!\!\!\!\!\!\!\!\!\!\!\!\!\!\! \langle I M_I|H_\textrm{sc}|I' M_I'\rangle = c_4  \delta_{I I'} \delta_{M_I M_I'} \nonumber \\
&&\!\!\!\!\!\!\!\!\!\!\!\!\!\!\!\! \times  \frac{1}{2} [I(I+1)-I_1(I_1 +1) - I_2 (I_2 + 1)].
\ea

We conclude this Appendix by presenting a convenient expression for dipole-dipole interaction between molecules $1$ and $2$ separated by $\mathbf{R} = (R, \theta', \phi')$, where $\theta'$ and $\phi'$ are the spherical angles of $\mathbf{R}$ in the $x$-$y$-$z$ coordinate system, which is defined with respect to the direction $\hat{\mathbf{z}}$ of the applied DC electric field. This expression is: 
\ba
H_\textrm{dd} &=& - \frac{\sqrt{6}}{4 \pi \epsilon_0 R^3} T^2(\mathbf{C}) \cdot T^2(\mathbf{d}^{(1)}, \mathbf{d}^{(2)}) \nonumber \\
&=& - \frac{\sqrt{6}}{4 \pi \epsilon_0 R^3} \sum_{p = -2}^{2} (-1)^p T_{-p}^2(\mathbf{C}) T^2_{p}(\mathbf{d}^{(1)}, \mathbf{d}^{(2)}), \label{eq:apdd1}
\ea
where $T_p^2(\mathbf{C}) = C^k_p(\theta', \phi') = \sqrt\frac{4 \pi}{2 k + 1} Y_{k,p}(\theta',\phi')$ and where we used Eq.\ (\ref{eq:sprod}). In the present manuscript, we only use the $p = 0$ component, for which [using Eq.\ (\ref{eq:tprod})]
\ba
 T^2_{0}(\mathbf{d}^{(1)}, \mathbf{d}^{(2)}) &=& \frac{1}{\sqrt{6}} \left(d^{(1)}_-  d^{(2)}_+ + 2 d^{(1)}_0  d^{(2)}_0 + d^{(1)}_+  d^{(2)}_-\right),\label{eq:apdd2}\\
 T_{0}^2(\mathbf{C}) &=& \frac{1}{2} (3 \cos^2 \theta' - 1). \label{eq:apdd3}
\ea
It is easy to see from Fig.\ \ref{fig:angles} that $\cos \theta' =\hat{\mathbf{R}} \cdot \hat{\mathbf{z}} = \sin \Theta_0 \cos(\Phi - \Phi_0)$.


\section{Interaction-Assisted Tunneling  \label{sec:iat}}

In this Appendix, we analyze the small corrections to the two approximations made to arrive at Eq.\ (\ref{eq:hdd1}). The two approximations were: (1) the extent of the Wannier function $w$ is much smaller than the distance between the sites, and (2) only terms involving two sites and conserving the number of molecules on each site contribute. Making the second approximation, corrections to the first approximation lead to the replacement of $V_\textrm{dd}(\mathbf{R_i} - \mathbf{R_j})$ in Eq.\ (\ref{eq:hdd1}), when $i$ and $j$ are close to each other, with a more complicated dependence on $i$, $j$ and on the internal state of the molecules at $i$ and $j$. However, even the separation of nearest neighbor sites ($\sim 500$ nm) is typically much larger than the extent of Wannier functions in a deep lattice ($\sim 50$ nm). 

Corrections to the second approximation result in interaction-assisted tunneling.  Interaction-assisted tunneling corresponds to the terms in Eq.\ (\ref{eq:hdd0}) where $j_4 = j_1$, while $j_2$ and $j_3$ are nearest neighbors not equal to $j_1$. Terms where $j_2$ and $j_3$ are not nearest neighbors are much smaller. Terms where $j_2 = j_3$ while $j_1$ and $j_4$ are nearest neighbors are identical to the case we are describing, giving an overall factor of 2. Another factor of 2 comes from the fact that $\langle j_2,j_3\rangle$ counts each nearest-neighbor pair only once. The resulting Hamiltonian is approximately equal to (the subscript in $H_\textrm{iat}$ stands for interaction-assisted tunneling)
\ba
&& \!\!\!\!\!\!\!\! H_\textrm{iat} = 2 \!\!\!\!\!\!\!\!  \sum_{\textrm{\begin{scriptsize}$\begin{array}{c}j_1 \sigma' \\ \langle j_2, j_3\rangle \!\! \neq \!\! j_1 \end{array}$\end{scriptsize}}} \!\!\!\!\!\!\!\!   \Bigg\{  \mu_{01}^2 \left[V_\perp(j_1, j_2,j_3) S^\dagger_{j_1} c^\dagger_{j_2 1 \sigma'} c_{j_3 0 \sigma'} + \textrm{h.c.}\right] \nonumber \\
&& \!\!\!\!\!\!\!\! + \sum_{m m'}  \mu_{m}  \mu_{m'} V_{m m'} (j_1, j_2,j_3) n_{j_1 m} c^\dagger_{j_2 m' \sigma'} c_{j_3 m' \sigma'}\Bigg\},\label{eq:iat}
\ea
where 
\ba
V_\perp(j_1, j_2,j_3) &=&  \int d^3 \mathbf{R} d^3 \mathbf{R'} V_\textrm{dd}(\mathbf{R}- \mathbf{R'})  \nonumber \\
&& \times w_{j_1 0}(\mathbf{R})  w_{j_1 1}(\mathbf{R}) w_{j_2 1}(\mathbf{R'}) w_{j_3 0} (\mathbf{R'}),\nonumber \\
V_{m m'} (j_1, j_2,j_3) &=&  \int d^3 \mathbf{R} d^3 \mathbf{R'} V_\textrm{dd}(\mathbf{R}- \mathbf{R'})  \nonumber \\
&&\times w^2_{j_1 m}(\mathbf{R}) w_{j_2 m'}(\mathbf{R'}) w_{j_3 m'} (\mathbf{R'}).
\ea
Physically, the interaction-assisted tunneling means that the presence of a molecule on site $j_1$ assists in a ``tunneling" of a molecule from site $j_3$ to site $j_2$. In the term proportional to $V_{mm'}$, this ``tunneling" does not change the internal states of the two molecules, while in the term proportional to $V_\perp$, it is accompanied by an exchange of a rotational excitation between the two molecules. 

Let us compare the magnitude of the tunneling amplitude $t$ to the magnitude of the interaction-assisted tunneling. In the deep-lattice limit, the tunneling amplitude $t$ in a 1D potential  $V_0 \sin^2(K X)$ (where $K = 2 \pi/\lambda$), is reduced relative to the recoil energy $E_R = \hbar^2 K^2/(2 M_m)$ by a factor porportional to $ \exp\left(-2 \sqrt{V_0/E_R}\right)$ \cite{bloch08}. 
Since interaction-assisted tunneling also involves an overlap of Wannier functions on neighboring sites, we may expect it to fall off similarly with increasing $V_0$.
At the same time, the reference energy scale for interaction-assisted tunneling is  $E_\textrm{dd}  = d^2/(4 \pi \epsilon_0 (\lambda/2)^3)$, the strength of dipole-dipole interaction between nearest-neighbor sites. For KRb with $\lambda = 1064$ nm \cite{demiranda10},  $E_R \approx (2 \pi) 1.4 \textrm{ kHz} >  E_\textrm{dd} \approx (2 \pi) 0.3$ kHz,
so we may expect the interaction-assisted tunneling to be smaller than the usual tunneling. 

\begin{figure}[b]
\begin{center}
\includegraphics[width = 0.8 \columnwidth]{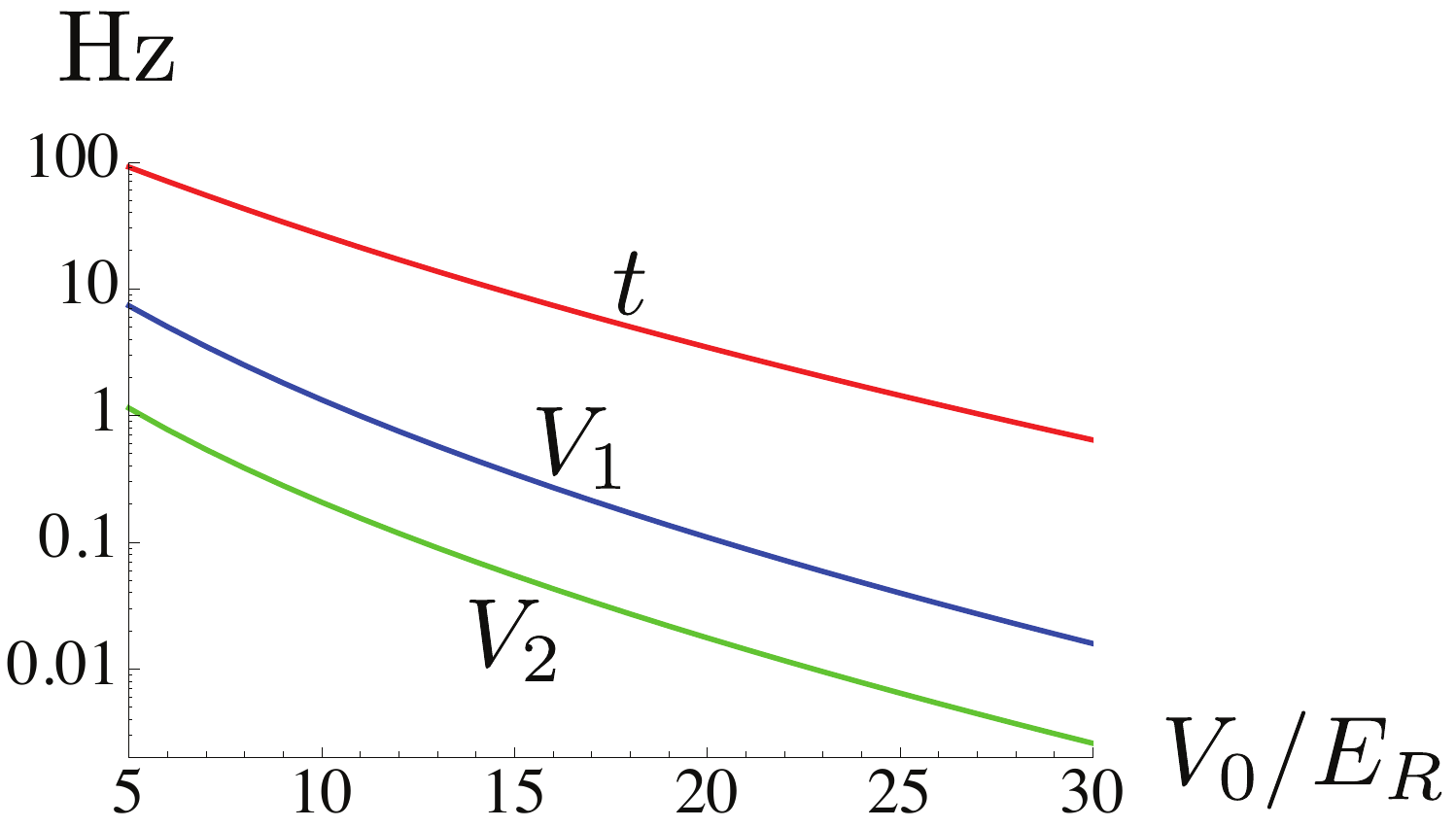}
\caption{(color online). The tunneling amplitude $t$ and interaction-assisted tunneling amplitudes $V_1$ and $V_2$ as a function of $V_0/E_R$, where $E_R$ is the recoil energy and where $V_0$ is the amplitude of the lattice. The vertical axis is in Hz. We use $\lambda = 1064$ nm and the mass and permanent dipole moment of KRb. \label{fig:iat}}
\end{center}
\end{figure}

To be more precise, in Fig.\ \ref{fig:iat}, 
we compare the magnitude of interaction-assisted tunneling to the usual tunneling. Let $w(X)$ be the 1D Wannier function for the potential $V_0 \sin^2(K X)$. For the case when the $(X,Y)$ coordinates of the three sites are (in units of $a = \lambda/2$) $j_1 = (0,0)$, $j_2 = (1,0)$, and $j_3 = (2,0)$, we estimate the amplitude of interaction-assisted tunneling as $V_1 = -\int_{a/2}^{5 a/2} d X \frac{d^2}{4 \pi \epsilon_0 X^3} w(X- a) w(X-2 a)$. We do not integrate from $X = -\infty$ to avoid integrating over the singularity of the $1/X^3$ potential at $X = 0$, which is unphysical and stems from the fact that $1/X^3$ interaction breaks down at small $X$. For the case $j_1 = (0,1)$, $j_2 = (0,0)$, and $j_3 = (1,0)$, we estimate the amplitude of interaction assisted tunneling as $V_2 = \int_{-\infty}^\infty d X \frac{d^2}{4 \pi \epsilon_0 (a^2 + X^2)^{3/2}} w(X) w(X-a)$. Solving numerically for $w(X)$ and for the tunneling amplitude $t$, assuming the mass and dipole moment of KRb and $\lambda = 1064$ nm,  in Fig.\ \ref{fig:iat}, we plot $V_1$, $V_2$, and $t$ as a function of $V_0/E_R$. We see that for the presented values of $V_0/E_R$, the interaction-assisted tunneling is at least 10 times weaker than the usual tunneling amplitude $t$, which confirms our expectations and allows to ignore interaction-assisted tunneling.

\bibliography{./refs}

\end{document}